%% file: main.tex
\documentclass[a4paper,fleqn]{cas-dc}

\usepackage[numbers]{natbib}

\usepackage[utf8]{inputenc}
\usepackage{graphicx}%
\usepackage{multirow}%
\usepackage{amsmath,amssymb,amsfonts}%
\usepackage{amsthm}%
\usepackage{mathrsfs}%
\usepackage[title]{appendix}%
\usepackage{xcolor}%
\usepackage{textcomp}%
\usepackage{manyfoot}%
\usepackage{booktabs}%
\usepackage{algorithm}%
\usepackage{algorithmicx}%
\usepackage{algpseudocode}%
\usepackage{listings}%

\graphicspath{{figure}}
\usepackage{subfig}
\usepackage{textcomp}
\usepackage{hyperref}
\usepackage{cleveref}
\usepackage[inline]{enumitem}

\usepackage[belowskip=-10pt,aboveskip=0pt]{caption}

\usepackage{todonotes}

\theoremstyle{definition}
\newtheorem{definition}{Definition}

\def\tsc#1{\csdef{#1}{\textsc{\lowercase{#1}}\xspace}}
\tsc{WGM}
\tsc{QE}

\newcommand{\orcidID}[1]{\href{https://orcid.org/#1}{\includegraphics[width=8pt]{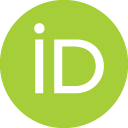}}}

\begin{document}
\let\WriteBookmarks\relax
\def\floatpagepagefraction{1}
\def\textpagefraction{.001}

\shorttitle{Semantic Data Management in Data Lakes}    

\shortauthors{S. Hoseini, J. Theissen-Lipp, C. Quix}  

\title [mode = title]{Semantic Data Management in Data Lakes}  



%

\author[1,2]{Sayed Hoseini}[orcid=0000-0002-4489-9025,
							bioid=1]

\cormark[1]

\fnmark[1]

\ead{sayed.hoseini@hs-niederrhein.de}



\affiliation[1]{organization={Department of Electrical Engineering and Computer Science, Hochschule Niederrhein University of Applied Sciences},
           addressline={Reinarzstrasse 49}, 
            city={Krefeld},
            postcode={47805}, 
            state={NRW},
            country={Germany}}

\author[2]{Johannes Theissen-Lipp}[orcid=0000-0002-2639-1949]





\affiliation[2]{organization={Chair of Databases and Information Systems, RWTH Aachen University},
	addressline={Ahornstraße 55, Aachen}, 
	city={Aachen},
	postcode={52074}, 
	state={NRW},
	country={Germany}}

\author[3]{Christoph Quix}[orcid=0000-0002-1698-4345]





\affiliation[3]{organization={Department of Data Science and Artificial Intelligence, Fraunhofer Institute for Applied Information Technology FIT},
	addressline={Schloss Birlinghoven}, 
	city={St. Augustin},
	postcode={53757}, 
	state={NRW},
	country={Germany}}

\cortext[1]{Corresponding author}



\begin{abstract}
In recent years, data lakes emerged as a way to manage large amounts of heterogeneous data for modern data analytics. One way to prevent data lakes from turning into inoperable data swamps is semantic data management. Some approaches propose the linkage of metadata to knowledge graphs based on the Linked Data principles to provide more meaning and semantics to the data in the lake. Such a semantic layer may be utilized not only for data management but also to tackle the problem of data integration from heterogeneous sources, in order to make data access more expressive and interoperable. In this survey, we review recent approaches with a specific focus on the application within data lake systems and scalability to Big Data. We classify the approaches into (i) basic semantic data management, (ii) semantic modeling approaches for enriching metadata in data lakes, and (iii) methods for ontology-based data access. In each category, we cover the main techniques and their background, and compare latest research. Finally, we point out challenges for future work in this research area, which needs a closer integration of Big Data and Semantic Web technologies.
\end{abstract}


\begin{highlights}
\item Definitions and Taxonomy: Provides clear definitions of terms related to Semantic Data Management and presents a taxonomy, classifications and comparisons of systems and approaches in this area.
\item Review of metadata models for data lake systems: Reviews metadata models for data lake systems that address semantics, defines requirements, derives evaluation criteria, and performs a detailed comparison.
\item Approaches for Semantic Modeling: Details algorithms for creating semantic models, distinguishing between semantic labeling and semantic modeling approaches..
\item Ontology-Based Data Access: Explores the use of semantic information for OBDA, where a Knowledge Graph serves as a common data model for specifying queries, and reviews scalable approaches for query processing over Big Data including a comparison and historical timeline.
\item Challenges in Semantic Data Management: Discusses the challenges that remain to be addressed in the field of semantic data management.
\end{highlights}

\begin{keywords}
 \sep Semantic Data Management \sep Semantic Web \sep Big Data \sep  Data Lakes \sep Ontology-based Data-Access 
\end{keywords}

\maketitle

\input{chapters/intro.tex}
\input{chapters/obdm}

\input{chapters/SM}
\input{chapters/obda}

\input{chapters/applications}

\input{chapters/challenges}

\input{chapters/concl}

\printcredits

\bibliographystyle{elsarticle-num}

\bibliography{main}

%

\end{document}

%% file: chapters/intro.tex
\section{Introduction}
Due to the growing dependence on data, its seamless provision and consumption becomes more and more important for different stakeholders. 
Data lakes have been proposed to tackle challenges in managing unstructured and structured data sources \cite{Hai2018}. 
Many organizations have established data lakes to collect data from heterogeneous sources,
which can then be used in data science projects \cite{pomp2018applying, bionda2019smart, Kharlamov2017a, yahya2021semantic}. 

The description of data sources with meaningful metadata is essential for the usability of data. 
Especially for users with limited domain knowledge, or when they are unfamiliar with a data set, meaningful metadata are indispensable. 
The mapping of raw data from data sources to semantically rich models increases the usability and interpretability of data.
Here, the usage of knowledge graphs (KGs) equipped with an expressive ontology to ensure well-defined meaning \cite{DBLP:journals/csur/HoganBCdMGKGNNN21, Ehrlinger2016} has proven useful in various ways, e.g., for data integration using ontology-based data access (OBDA) \cite{DBLP:conf/ijcai/XiaoCKLPRZ18}.

These advantages have been recognized in data lake research and practice in the recent years. 
While first implementations aimed at processing Big Data efficiently using distributed, scalable systems (such as Hadoop), the need for proper management of metadata and data quality has also been recognized \cite{constance,Farid2016}. 
Several approaches have been proposed for semantic data management (SDM) in data lakes.
Semantic data lakes need to integrate Big Data and Semantic Web technologies. 
In the Semantic Web, the Resource Description Framework (RDF) and the Web Ontology Language (OWL) are the main languages to represent data as a set of linked data items \cite{DBLP:books/igi/11/BizerHB11}. 
These languages can be used also to represent metadata in form of ontologies or KGs.
Although the technologies for linked data and the Semantic Web have become more mature in the recent years, the amount of data considered in Semantic Web applications is far less than in Big Data applications. 
Thus, scalability to large, heterogeneous data sets is major challenge for applying Semantic Web technologies in data lakes.

\begin{figure*}[t]
    \centering
    \includegraphics[height=0.8\columnwidth]{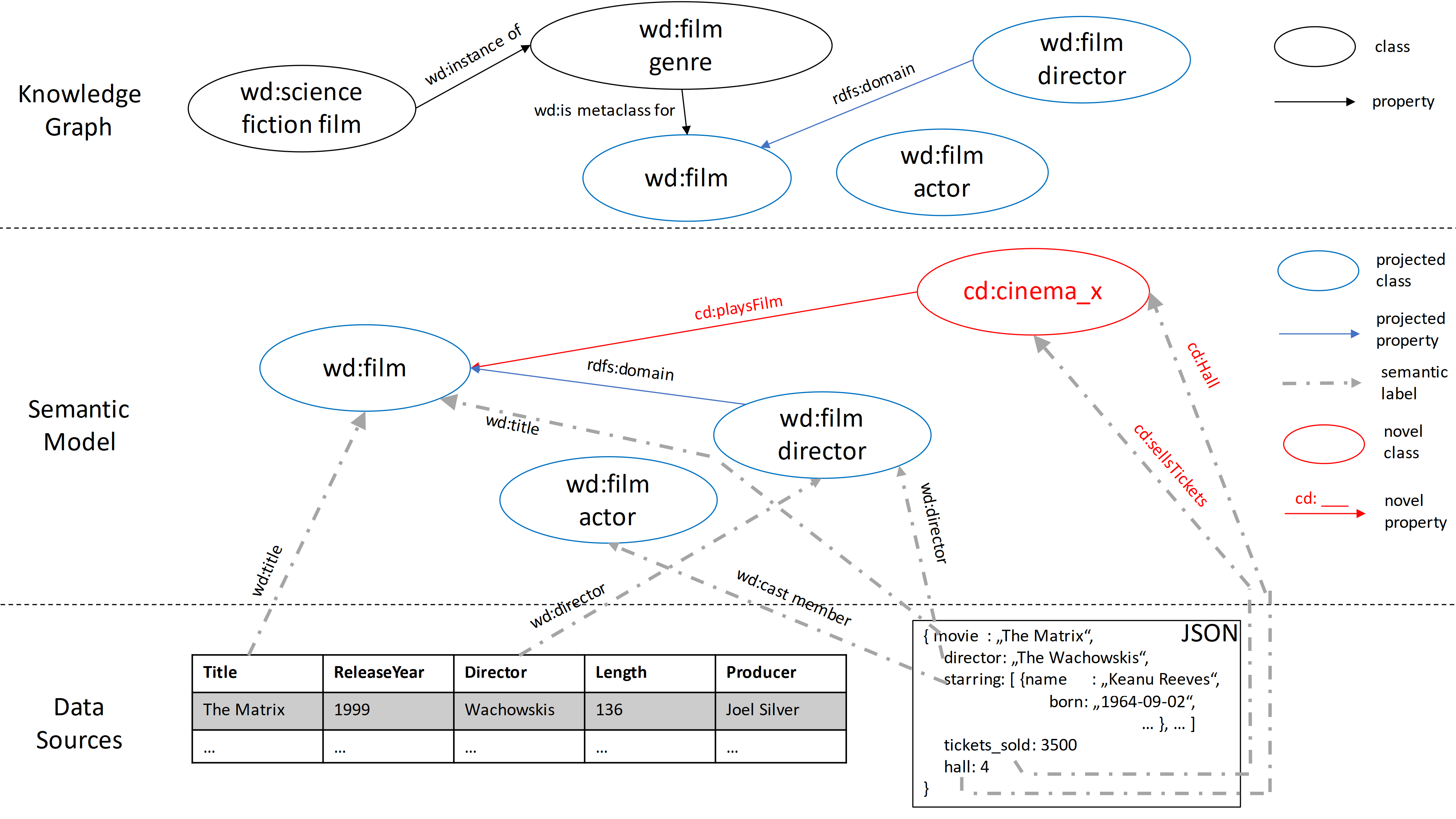}
    \caption{Semantic data management for heterogeneous data sources in a data lake. 
    Most concepts of the semantic model are drawn from the KG, but in order to describe specific data sets one may define novel classes and relationships.}
    \label{fig-intro-overview}
\end{figure*}

Semantic data extends the regular, extractable metadata (e.g., schema, data types) to convey context information that is not inherent to the individual data source at hand. 
According to \cite{Paulus2021}, there are two important processes that ensure the quality and acceptance of SDM: (1) \textbf{the creation and maintenance of conceptualizations} as well as (2) \textbf{the description of data sources in the form of semantic models}, i.e., linking source schemas to the conceptualization.
A semantically well-annotated data source can be discovered using the conceptual representations of the data and understood using the provided context information stored in the model. 
The creation of such semantic models involves decoding the current data source, checking suitable KGs, and linking data attributes to concepts in the KG. 
The KG can therefore be seen as a common data model which provides a conceptual description of the data resources of an organization. As companies recognize the increasing value of data for their business processes, such models are frequently being established as part of a data governance strategy or a master data management program \cite{Dibowski2020a}.
The modeling of a KG is a complex process which requires human intelligence; in this article, we focus on the mapping between data sources and the KG. We assume that the KG is already available; it might be extended during semantic modeling to capture the complete semantics of the data sources.

The semantic model can be viewed as a \emph{projection} of the entities and relationships of the KG onto the data sources. 
The semantic model is an additional layer between the data layer and the knowledge layer \cite{pomp2018applying}.
The basic idea of the semantic model is sketched in \Cref{fig-intro-overview}. 
The raw data sets in the data lake are represented in the lower data layer; they can be in different formats, such as tabular data or hierarchical JSON data, but have partially overlapping content. 
Typically, data lakes are able to extract schema information from the sources automatically.
The semantic model can be represented as a graph with nodes and edges derived from the KG, enriched with nodes and edges to represent the mapping and the extended semantics of the data sources.
In the example of \Cref{fig-intro-overview}, we reuse concepts and properties from the WikiData ontology (prefix \textit{wd}) as a part of our KG.
By explicitly encoding the semantic type and relationship between source attributes in the graph, the semantic model precisely describes the intended meaning of the data source. 
A SDM platform should support the process, e.g., by proposing appropriate concepts and related mappings when creating the semantic model.

Furthermore, instead of only using a static KG, the semantic model may also introduce novel classes and properties.
This need for evolution arises when users provide data sources that include concepts and relationships which are not covered by the KG.
In the example, the JSON document contains data about a specific cinema (prefix \textit{cd} (cinema domain)) that needs to be modeled explicitly.
This novel knowledge needs to be incorporated in a structured way, thereby continuously evolving the knowledge layer \cite{Pomp2019}.

\emph{Semantic labels} are the direct annotations of elements of the data source (e.g., schema attributes) with elements of the semantic model. The semantics are defined by a suitable predicate and object. A semantic label is usually represented as a triple: (schema attribute, predicate, object). 
In \Cref{fig-intro-overview}, the semantic label of the table's attribute ‘‘Title’’ is constructed through the object ‘‘wd:film’’ and the predicate ‘‘wd:title’’ modeling the relationship between them.
A semantic data lake should assist the extraction of a semantic model from the raw data sets, i.e., a more conceptual description of data sets including concepts and their relationships. 
The semantic models are then linked to the data sets to know which data set is related to which concept \cite{Dibowski2020a}.

\begin{figure}[tb]
\centering
  \includegraphics[width=0.95\columnwidth]{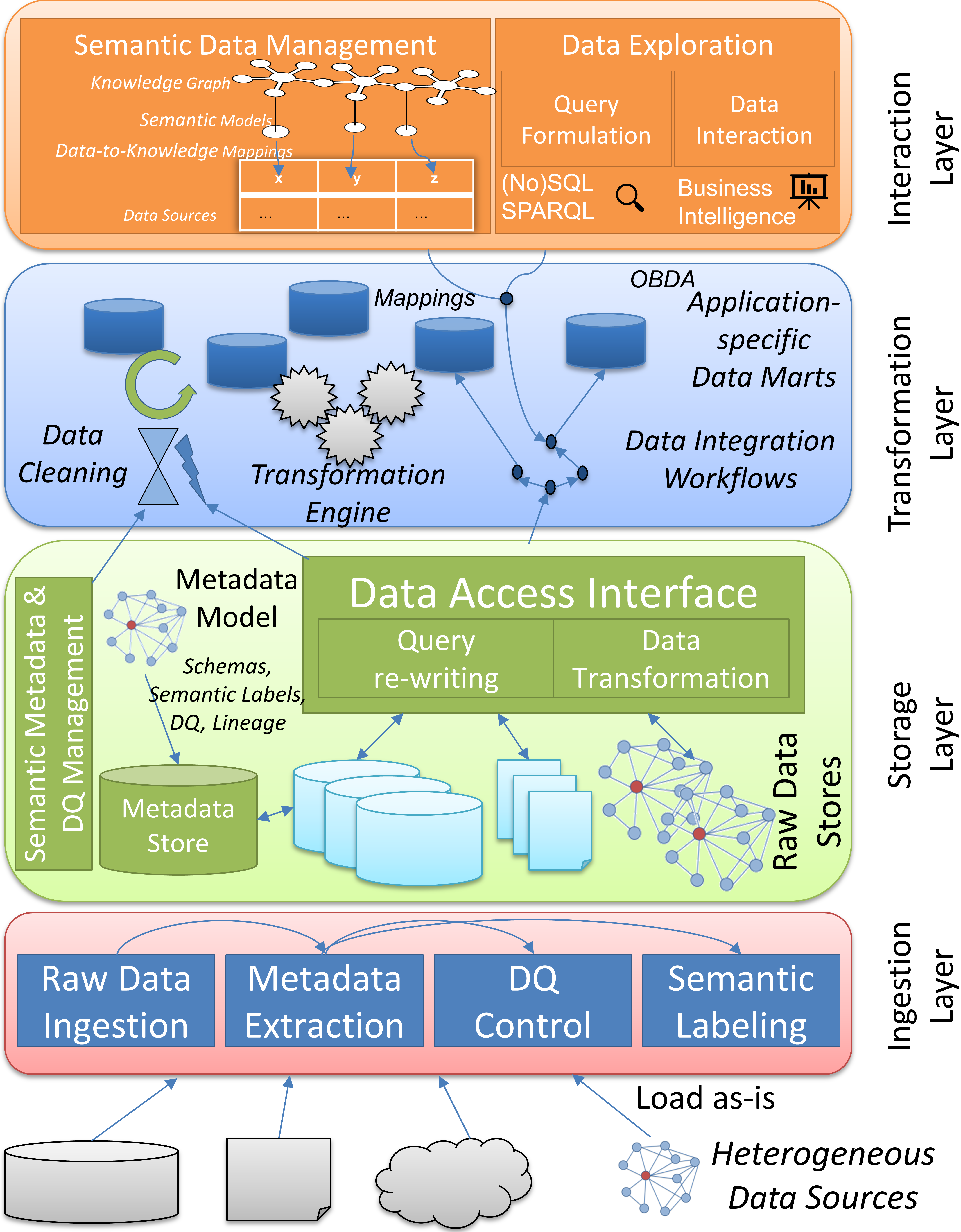}
  \caption{Semantic data lake architecture, extended from \cite{Quix2019}}
  \label{fig-data-lake-arch}
\end{figure}

In \Cref{fig-data-lake-arch}, we propose an extended version of the data lake architecture in \cite{Quix2019}, where especially metadata-related functions are enriched with semantics. 
For example, a semantic labeling component in the ingestion layer adds semantic labels to the extracted metadata elements. 
Data quality (DQ) management can be also improved, e.g., by verifying whether semantic information is present for a data source. 
The semantic information (labels, models, KG, etc.) is managed in the storage layer in an extend semantic metadata repository, which provides, for example, the mappings required for OBDA. 
In particular, the semantic metadata should facilitate the usage and interpretation of data.
Therefore, the interaction layer has several additional components, e.g., for browsing the KG and semantic models, for defining queries using a semantic query language such as SPARQL\footnote{\url{https://www.w3.org/TR/rdf-sparql-query/}}, and editors for refining the semantic mappings and models.

\subsection{Related surveys, research methodology and contribution}
In this survey, we give an overview of the recent developments in SDM related to \textbf{semantic} data lakes in particular. In contrast to other surveys in this field, discussed briefly in the following, we focus especially on approaches which can be \textbf{scaled up to Big Data} and facilitate the \textbf{semantic integration of data}. 

Sawadogo et al. \cite{Sawadogo2021} focus on generic data lake architectures and metadata management and
discuss the pros and cons of data lakes and their design alternatives.
Similarly, Hai et al. \cite{DBLP:journals/corr/abs-2106-09592} present a comprehensive overview of research questions for designing and building general data lakes. They classify the existing approaches and systems based on their provided functions for data lakes and provide a thorough comparison of existing solutions and the discussion of open research challenges.

Paulus et al. \cite{Paulus2021} investigate existing semantic modeling approaches, discuss their strengths and weaknesses for real-world use and present future challenges and necessary research directions that the community needs to focus on in order to make semantic data management acceptable in everyday business. Data lakes and scalability to Big Data is not in the focus of Paulus et al. 

Xiao et al. present first a more theoretical discussion on the formal implementation details in OBDA \cite{DBLP:journals/dint/XiaoDCC19}. In a follow-up work \cite{DBLP:conf/ijcai/XiaoCKLPRZ18}, they present the tooling ecosystem and concrete use cases in a wide range of commercial applications. Additional works in this context and a more detailed discussion can be found in \Cref{chapter-OBDA}. 

\begin{figure*}[h]
	\centering
	\includegraphics[width=1.5\columnwidth]{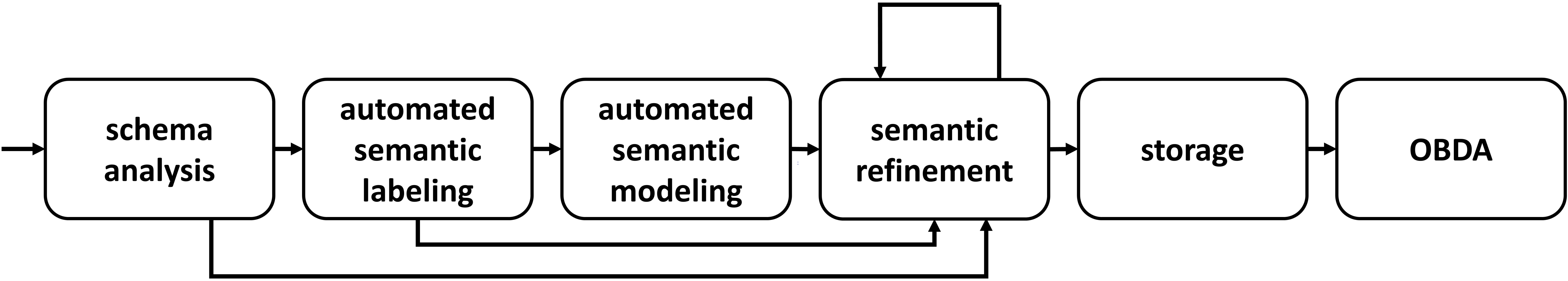}
	\caption{Overview of the phases during the automated generation of a semantic model as illustrated in \cite{Paulus2021a} extended by OBDA}
	\label{fig-DL-PLASMA}
\end{figure*}

A recent survey by Liu et al. \cite{liu2022tabular} reviews approaches for semantic labeling (see \Cref{sec:semantic-labeling}). It is closely related to this work, but we take a more holistic approach. We address the complete process, starting from SDM in general in semantic data lakes, to semantic modeling for data integration and finally OBDA. In each section, we focus especially on the applicability to data lakes and Big Data. To provide a clear and logical presentation, we were inspired by the pipeline presented by \cite{Paulus2021a} (see \Cref{fig-DL-PLASMA}), which has been extended by OBDA, because it provides a uniform data access interface for heterogeneous data in data lakes.

To collect the relevant papers, we made use of Google Scholar primarily.
We first used a list of keywords, which was manually created when reading relevant papers starting from e.g. \cite{Paulus2021, Dibowski2020a, Mami2019}) and evolved dynamically during the course of the writing:
`semantic data lake', 
`knowledge graph data lake',
`semantic data catalog',
`semantic data integration',
`semantic / ontology-based data management',
`semantic models',
`semantic annotation',
`semantic table interpretation',
`semantic type detection',
`ontology-based-data-access'.
For each query, we first retrieve relevant results by importing the DOI into \textit{Citavi}, a program for reference management, then read each retrieved paper to judge whether the paper is relevant for one of the three domains of interest (Semantic Data Lakes, Semantic Data Management / Integration, OBDA). 
From there we assigned each paper to a category resembling the structure of this article.
We made heavy use of the \textit{citation snowballing method} \cite{10.1145/3340531.3417426}, i.e., extending the results by searching through the cited references in the initial keyword-based findings as well as exploiting the `Cited By' functionality in Google Scholar.
The commercial systems were found using Google Search.

In summary, this article makes the following contributions:
\begin{itemize}[leftmargin=0.5cm]
    \item A definition of the terms related to SDM is provided in \cref{sec:defintions}. Terms like semantic models or
    semantic labels are frequently used in the context of SDM; in order to make a proper classification and comparison of systems and approaches in this area, we first need clear definitions. Additionally, a taxonomy (\Cref{fig-sdm-taxonomy}) of the various notions is presented as a high-level overview.
    \item Following the basic definitions for SDM, we review metadata models and data lakes addressing semantics in some way in sections \ref{sec:semantic-labeling} to \ref{chapter-compare-datalakes}. We define requirements for metadata models and data lakes, derive criteria for evaluation, and perform a detailed comparison.
    \item \Cref{chapter-automatic} addresses in detail approaches for semantic modeling, i.e., algorithms for creating the semantic models (as sketched in \Cref{fig-intro-overview}) (semi-)automatically. We distinguish between approaches for semantic labeling and semantic modeling. 
    \item The most advanced use of semantic information is for ontology-based data access (OBDA). A KG is used as common data model for specifying queries; OBDA systems take mappings to data sources into account and translate queries into the query language of the particular data source. Especially in this context, scalable approaches are required for query processing over Big Data. We review several systems in this area that have been proposed in the recent years (\cref{chapter-OBDA}).
    \item In \Cref{chapter-applications}, we discuss two application areas for SDM in data lakes: Industrial Internet-of-Things and Smart Cities. This shows that the SDM techniques can be applied to real-world use cases.
    \item Finally, we derive in \cref{sec:challenges} challenges that still need to be addressed in the field of semantic data management.
\end{itemize}

%% file: chapters/obdm.tex
\section{Semantic Data Management in Data Lakes} 
\label{chapter-OBDM}
We will first provide definitions for the main terms used in this survey (\cref{sec:defintions}) and discuss in detail the distinction between semantic labeling and semantic modeling (\cref{sec:semantic-labeling}). The subsequent sections then review the state of the art for metadata models for data lakes with semantics (\cref{sec:metadata-models}), semantic data lake systems (\cref{chapter-DL-systems}), and commercial systems (\cref{sec:commercial-systems}). The final subsection summarizes the discussion and provides a detailed comparison.

\subsection{Definitions}
\label{sec:defintions}
We have been using terms like semantic data lakes, semantic labels, and semantic models in the introduction and now clarify their meaning.
All given definitions should be general and are not limited to a specific technology. Semantic tools often apply W3C standards (e.g., RDF, OWL), but some industrial solutions do not focus on compatibility with a standard but model KGs and semantic metadata in a different way.
\begin{definition}\label{definition-datalake}[\textbf{Data lake}]
``\textit{A data lake is a flexible, scalable data storage and management system, which ingests and stores raw data from heterogeneous sources in their original format, and provides query processing and data analytics in an on-the-fly manner.}'' \cite{DBLP:journals/corr/abs-2106-09592}
\end{definition}
In particular, here query processing in an on-the-fly-manner refers to the ELT (Extract-Load-Transform) principle \cite{Quix2019}. 
In data warehouses, the classical ETL (Extract-Transform-Load) idea is applied: data is transformed into a uniform schema before it is loaded. 
Only then, queries can be executed on this integrated data set. 
In data lakes, raw data is stored in a heterogeneous storage layer, query processing needs to be able to work with heterogeneous storage systems (e.g., as Apache Spark does with its query language SparkSQL). 
In the Big Data context, this behavior is also referred to as schema-on-read.
This is also an important aspect for the scope of this survey: we only consider approaches that do not require the data to be converted into a uniform schema, as such approaches are just not scalable to Big Data.

\begin{definition}\label{definition-semanticdatalake}[\textbf{Semantic data lake}]
Semantic data lakes are a specific form of traditional data lakes which extend the capabilities through a \textit{semantic layer} that enriches and connects the stored data \textit{semantically}. 
The \textit{semantic layer} equips data sets in the lake with connections from the data set's metadata to conceptual/logical models, which encapsulate knowledge potentially external to the content of the data, such as domain knowledge. 
\end{definition}

\begin{definition}\label{definition-semanticlabels}[\textbf{Semantic labels}]
Semantic labels manifest 
the connection between data sources and concepts of a conceptualization with the objective to describe the semantics of the data source's attribute to which the label is associated to.
\end{definition}

Semantic labels are the core concept of implementing semantic data management in practice.
They complement extractable metadata (such as data types, sizes, formats, etc.) to convey \textbf{context} information which may not be inherent to the data source at hand.
The most common semantic labels are those which connect elements of a data source's schema in a One-to-one-relationship, e.g., the column of a table or a leaf node in a hierarchical/nested data set (e.g., JSON file), to their counterparts from a KG as previously described. 
However, they can also be used for example on the level of individual data values, for a specific row of the table, to denote relationships between individual schema entries, the entire data set, or any other relevant concept.
In the literature, various terms are used for semantic labels, for example, semantic enrichment, links, tags, types, profiling, or annotations \cite{Scholly2021, Zhang2020, Quix2016, Sawadogo2021}. 
The idea behind semantic labels differs from approaches that model their entire metadata/data catalog as a KG, such as \textit{Aurum} \cite{Fernandez2018a}, 
by the fact that their primary objective is to provide context, hence additional information from external sources, in contrast to the latter which store metadata in a graph-based data model.

In the area of OBDA (see \cref{chapter-OBDA}), the term mapping is also commonly used. 
Mappings establish a connection between data attributes and KGs and they are the technical prerequisite for query rewriting, e.g., from SPARQL to SQL.
While mappings in OBDA refer to different notions technically, on a conceptual level however, they can be regarded as closely related to semantic labels, because essentially they connect the metadata of a data source to instances of a conceptualization. The main difference is that mappings in OBDA require a precise semantics in order to rewrite queries correctly. Semantic labels are usually not defined with precise semantics in mind, as they aim at providing more context information to the user or at connecting related data sets.

\begin{definition}\label{definition-semanticmodels}[\textbf{Semantic models}]
A semantic model specifies the broader context of a data source by extending the semantic labels through additional meaningful concepts and provides suitable relationships that hold between various concepts. 
\end{definition}
\begin{figure*}[H]
	\centering
	\includegraphics[width=0.65\textwidth]{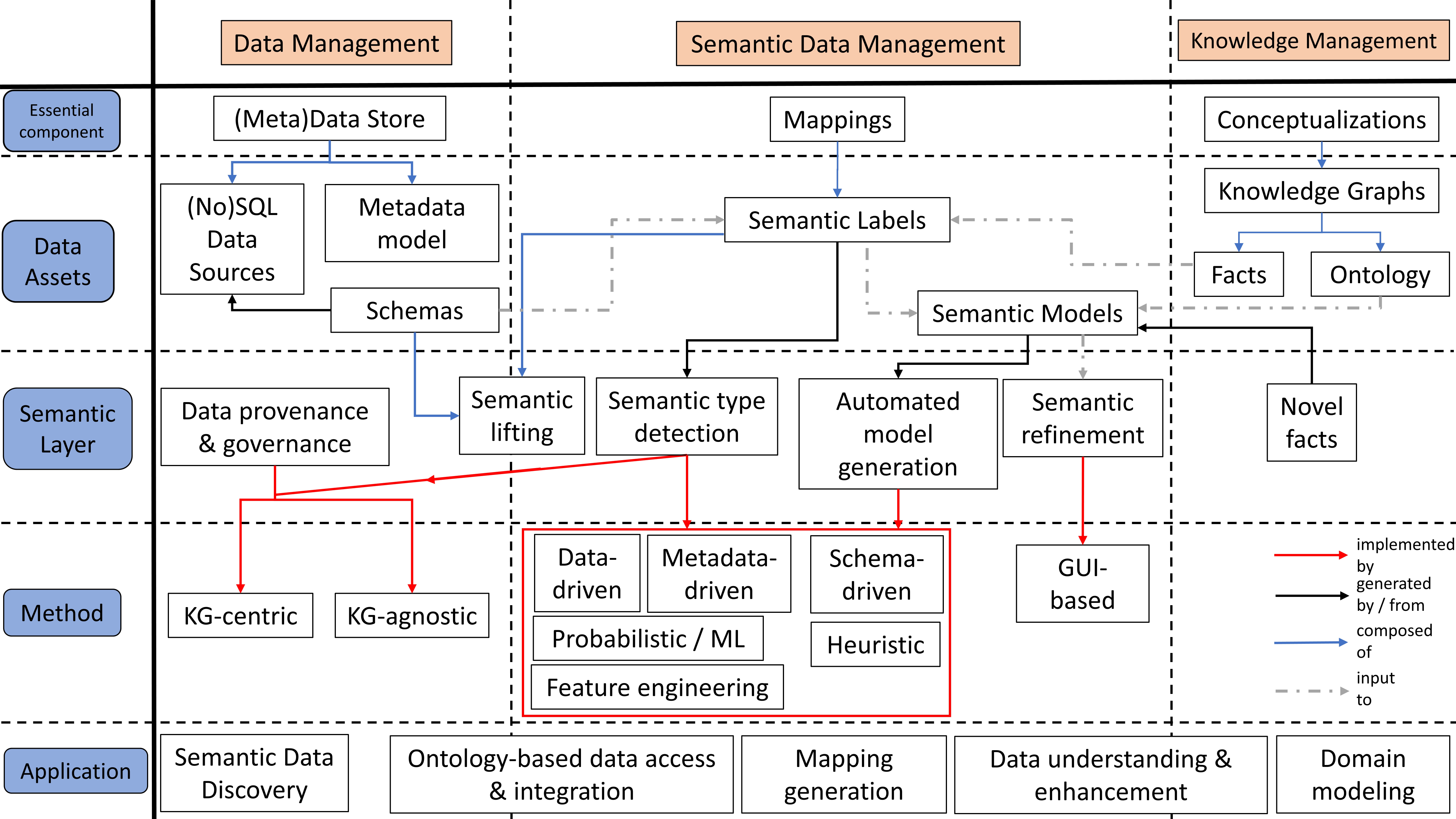}
	\caption{A Taxonomy for the concepts and methods discussed in this survey}
	\label{fig-sdm-taxonomy}
\end{figure*}

To make the definitions more clear, we provide a rough formalization, explained by referring to the case study of \Cref{fig-intro-overview}. 
We consider two data sources stored in a data lake, a table and a JSON file where each data source $D$ consists of a set of attributes $D = (c_1, ..., c_n)$ (i.e., columns in a table or paths in hierarchical data sets, respectively). 
For the conceptualization, we consider a standard knowledge graph $KG$ defined by a set of RDF triples in the form of subject-predicate-object: $(s, p, o)$. 

The idea behind a semantic label for $c_i$ is to best describe the semantics of $c_i$ by defining a triple $SL_{c_i} = (c_i , p , o)$, with $p$ and $o$ preferably, but not necessarily, coming from $KG$. 
For example, the semantic label for the JSON key ``tickets\_sold'' is represented as the triple: 
(tickets\_sold, cd:sellsTickets, cd:cinema\_x).
Here, $p$ and the $o$ are novel concepts for a specific domain, that are not present in this form in the general-purpose ontology of WikiData. 
Users may incorporate these novel facts into the overall knowledge layer, thereby evolving it dynamically. 
On the other hand, the table's column ``Title'' and the JSON's key ``movie'' possess the same predicate and object and therefore provide a suitable entry point for data integration between the two heterogeneous data sets.

The semantic model $SM$ is a set of triples built upon the set of semantic labels $SL = \{SL_{c_1}, \ldots, SL_{c_n} \}$. In order to specify the broader context of the data source $D$, it may additionally contain further nodes and edges drawn from $KG$ denoted as $SM_{KG}$. 
The semantic model can be extend by concepts and relationships of the sources, which are not represented in $KG$. We denote
this extension of the semantic model as $SM_X$.
Note that $SL$, $SM_{KG}$, and $SM_X$ are pairwise disjoint.
\begin{center}
        $SM = SL \cup SM_{KG} \cup SM_X$ \newline
  (cd:cinema\_x, cd:playsFilm, wd:film) $\in SM_X$ \newline
  (wd:film director, rdfs:domain, wd:film) $\in SM_{KG}$ \newline
  (''Title'', wd:title, wd:Film) $\in SL$
\end{center}
The given case study is restricted to the schema of the two data sources for the purpose of illustration, but the same formalization can be generalized to various scenarios. 
Naturally, it can easily be applied to other concepts in the data layer, e.g., entire data sets and relationships between them (replace $c_i$ with data set $d_i$).
In \Cref{fig-intro-overview}, modeling the film genre of the movie ``The Matrix'' on the level of individual data entries (single row/object or specific cell entry/value in the table/JSON) of the data sources could have been included.   

Some approaches, like \textit{DoDuo} \cite{DBLP:conf/sigmod/SuharaL0ZDCT22}, establish the relationship between any two columns in a table directly via a predicate without an intermediate element of the semantic model in between. 
For example, the triple is composed of two table columns and only the predicate comes from the conceptualization: $(c_i, p , c_j)$.
Semantic labels, especially when considered as `weak' annotations, are frequently also modeled as just an additional field in the metadata repository and not expressed as a triple in RDF, e.g., in \textit{GEMMS} \cite{Quix2016}. 
If one thinks of the predicate being \href{https://schema.org/isRelatedTo}{isRelatedTo}, the given formalization still applies with no further specification.

\subsection{Semantic Data Management Taxonomy}\label{sec:semantic-labeling}

In \Cref{fig-sdm-taxonomy}, we have structured the various concepts and methods discussed in this survey into a taxonomy. 
The approaches are horizontally separated by the three main areas addressed in this article: data management, knowledge management and their particular intersection SDM.
We further differentiate five dimensions vertically which gradually decrease the level of abstraction to culminate into concrete applications which are described throughout the article. 
The first dimension represents higher-level conceptual components; in this article, these include a data lake system (i.e., (No)SQL sources and their metadata), semantic labels and models, as well as serialized knowledge in the form of KGs in the second dimension.
In the third dimension the semantic layer includes functionality that a semantic data lake system may implement to generate the semantics required for effective SDM, which are implemented by different methods in fourth dimension.

\textit{Semantic labeling}  is the process of creating semantic labels for a data source.
In the literature, the automation of this process is also denoted as \textit{Semantic Type Detection} (STD) \cite{Hulsebos2019} or \textit{Semantic Table Interpretation} (STI) \cite{liu2022tabular}.
Burgdorf et al.~\cite{Burgdorf2022} identify three different directions for this task:
\textit{schema-driven}, i.e., using the schema of each data point within a data set, such as \cite{Pinkel2017, Paulus2018};
\textit{data-driven}, i.e., using the actual data values contained in a data set to assign fitting concepts with statistics or machine-learning-based classification, such as \cite{Ramnandan2015, Pham2016, Abdelmageed2021a};  
\textit{metadata-driven}, i.e., using all available additional pieces of information on a data set that might contribute to the semantic labeling of the data, e.g., comments or textual data documentation, such as \cite{DBLP:journals/sigmod/DengSLWY22}.
Liu et al. divide approaches further into ML-based, heuristic or feature-engineering methods \cite{liu2022tabular}.

From the taxonomy it is evident that the concept of the semantic label, combining facts stored in the KG and schema attributes of the data sources, is central to the whole framework. This is one of the reasons why STDs have gained so much attention recently \cite{liu2022tabular}.
Another crucial property is whether the detection can be performed against arbitrary KGs (see \Cref{chapter-automatic}). 
The labels are the prerequisite for the generation of semantic models that are ultimately refined using a graphical interfaces and potentially contribute novel facts to the KGs.
Data provenance and governance can be enhanced or built entirely from these semantic data structures and again there exists an important distinction between KG-centric and agnostic approaches (see \Cref{chapter-compare-datalakes}).
Finally, there exists a variety of applications, i.e., semantic data discovery (e.g. semantic similarity \cite{10.1145/3340531.3417426}) or ontology-based data access (see \Cref{chapter-OBDA}). 

The Semantic Web Challenge on Tabular Data to Knowledge Graph Matching\footnote{\url{https://www.cs.ox.ac.uk/isg/challenges/sem-tab/index.html}} (\textit{SemTab}) is an annual competition (since 2019) for researchers to present novel approaches evaluated with regards to accuracy and usability on proposed benchmarks. The challenge's problems include three tasks to annotate a table with semantics. 

Cell Entity Annotation is a cell-based annotation task which deals with annotating the values in the table cells with entities inside a KG. Column Type Annotation (CTA) is a schema-level annotation task which aims to map the underlying table schema to a KG. Finally, Column Predicate Annotation (CPA) is a schema-level annotation that defines binary relationships (properties of the KG) between pairs of columns of the table.
\cite{liu2022tabular} further consider the tasks of Row-to-instance (R2I), i.e. annotating an entire row of a relational table with a KG entity and Topic annotation, i.e. annotating the entire table with a concept or an entity from the target KG.
Because Semantic Models, OBDA as well as most data integration techniques predominantly live on the level of a data sources schema, the only relevant tasks we consider here are CTA \& CPA.

With \textit{semantic modeling}, we refer to the process of creating relationships between the semantic labels thereby describing the data source's context.
In practice, semantic modeling often depends on accurate semantic labels, which can be viewed as elements of a semantic model of only the first order.
In \cref{chapter-automatic} we discuss recent papers that present algorithms to automate these processes.

Paulus et al.~\cite{Paulus2021a} introduce a phase called \textit{semantic refinement}. 
They reinforce the view that the automated creation of semantic labels and models is a self-evident requirement for any OBDM platform.
While automated approaches can provide initial semantic labeling and modeling after analyzing the schemata of the data sources, erroneously chosen or missing labels for concepts will have a severe impact on the semantic modeling phase. 
Therefore, the human operator shall have the opportunity to improve the quality of the semantic labels as well as the resulting model manually by checking, validating, correcting, selecting, or exchanging ambiguous or mutually exclusive concepts, as well as adding new concepts that were not included initially.

Mami et al.~\cite{Mami2016} describe a transformation called \textit{semantic lifting}.
Given a set of semantic labels and a (yet non-semantic) data set, a semantic lifting function returns a semantically-annotated data set with semantic labels for entities and attributes.
Mami et al. considered semantic labels only in their original proposal, however \textit{semantic lifting} can potentially be applied to include entire semantic models as well any other descriptive metadata (e.g., comments or textual data documentations stored in documentation tools or Wikis).
Transforming data sources to incorporate semantic information along with the raw values beforehand to prepare them for subsequent semantic access necessarily corresponds to an ETL approach contradicting the ELT principle that is inherent to data lakes.
When dealing with large data sets, this transformation can be challenging and expensive, particularly when data liveliness is important.

\subsection{Metadata models for data lakes that account for semantics}
\label{sec:metadata-models}
Metadata models for data lakes have been an active area of research in recent years and many existing proposals consider semantics as a fundamental issue for the management of heterogeneous data sources.

\textit{MEDAL} and its successor \textit{goldMEDAL}, \cite{Scholly2021} identify requirements for a generic metadata model for data lakes and consider semantic enrichment as a core requirement. 
The earlier proposals \textit{HANDLE} \cite{DBLP:journals/dke/EichlerGGSM21} as well as \textit{GEMMS} \cite{Quix2016} enable manual semantic labeling already. 

Other approaches model their metadata directly in the RDF language as an ontology. An early approach is presented by Farid et al.~\cite{Farid2016}, where all source data and metadata are stored in the RDF format. 
Diamantini et al.~\cite{Diamantini2018} choose a network-based and semantics-driven representation for their model that can be extended with external KGs. 
In a follow-up work \cite{Diamantini2021}, they describe another ontology-based metadata model tailored to query-driven analytics. Here, a special focus is laid on key performance indicators and statistical measures typical in data science, where the user specify the indicators of interest at query time, as well as related dimensions of analysis, in terms of KG concepts. 
The framework then performs the required data discovery and on-demand integration to provide the requested indicators.

A similar approach is followed by Bagozi et al.~\cite{Bagozi2019} who utilize indicators to enable personalised data lake exploration. 
Indicators are usually computed based on the centralization of the data storage, according to a less flexible ETL approach than what traditional data lakes offer. 
In addition, domain experts, who know the data stored within the data lake, are usually distinct from data analysts, who define indicators, and users, who exploit indicators. 
Hence, the authors allow domain experts to enrich the heterogeneous data sources within a data lake with semantic models using domain ontologies and further propose an ontology used by analysts to define indicators and analysis dimensions, in terms of concepts within semantic models as well as formulas to aggregate them. 
In \cite{Bianchini2018} and \cite{Bianchini2020}, Bianchini et al. demonstrate the effectiveness of this model in the Smart City environment extended by theoretical considerations about modeling user preferences to enable personalized exploration of data.

At Bosch, the potential of semantic technologies has already been exploited at scale in production. 
The development of KGs required an immense initial effort by the company, but promises a long-term solution for a series of sophisticated challenges. 
Dibowski et al.~\cite{DBLP:conf/semweb/KalayciGLXMKC20, Dibowski2020a} describe a metadata model represented as an ontology called \textit{DCPAC}, based on W3C recommendations such as the Data Catalogue Vocabulary (DCAT\footnote{\url{https://www.w3.org/TR/vocab-dcat-2/}}) and the PROV Ontology (PROV-O\footnote{\url{https://www.w3.org/TR/prov-o/}}) for data catalog management and provenance control.
Ingested data assets are enriched with semantics by aligning, annotating, and enriching the input data with DCPAC concepts. 
In addition, they utilize a new concept called ontology-driven, self-adaptive frontends, in which the data catalog's GUI can dynamically render information from a KG. A similar concept was presented in \cite{Calvanese2021}. 
This is useful because changes in the underlying ontology do not require any changes at the frontend.

\subsection{Existing semantic data lake systems}
\label{chapter-DL-systems}
There exist several proposals for data lake systems that include functions for SDM.

Google Dataset Search (Goods) \cite{DBLP:conf/sigmod/HalevyKNOPRW16} bootstraps its data catalog by crawling Google’s storage systems and logs in order to discover what data sets exist. 
It examines their content and metadata and matches it against Google’s Knowledge Graph (GKG) to identify entities and this may perhaps be the most obvious demonstration of the power semantics for data management.

Besides its metadata model, \textit{GEMMS} \cite{Quix2016} is a tool for the automatic metadata extraction in data lakes. 
The authors recognized that often acronyms or synonyms are used as the name of a schema element. By using semantic annotations, such ambiguities can be avoided and the semantics of elements can be clearly defined.
\textit{Constance} \cite{constance} takes up this idea and is one of the first data lake systems to enable semantic labels on the level of the data source's schema.

\textit{Apache ATLAS} is a stand-alone data governance framework for Hadoop that has been successfully integrated into data lakes \cite{Liu2020} allowing semantic labeling. 
\textit{goldMedal}  demonstrates their metamodel for data lakes using \textit{Apache ATLAS}. Hence, it is neither viewed as a standalone data lake system nor a metamodel, but may be utilized as a core technology worth of being mentioned.

The data discovery system \textit{Aurum} \cite{Fernandez2018a} has KGs as a central element. The system builds a so-called enterprise KG in a single pass through the data sources by using data summarization and hashing to capture relationships between them. \textit{Aurum} describes the data discovery problem by properties and relationships of the data sources stored in the data lake. Properties include schema labels and quality measures, relationships may include similarity and primary/foreign key relationships. 
In the KG, each node represents columns of the data sources, edges represent relationships between two nodes, and hyperedges connect any number of nodes that are hierarchically related, such as columns of the same table, or tables of the same source. 
\textit{Aurum} allows individual entries to be represented, but suggests that the additional storage costs are not offset by the gain in expressiveness.
To calculate the weights associated with each property, it considers content similarity (the values of the columns are similar), schema similarity (the attribute names are similar), or the existence of a foreign-key-relationship between them. As such, the underlying algorithm to generate the hypergraph involves two steps: firstly, each data source is profiled, and secondly, relationships are calculated based on the profiles. Querying is performed through a language called Source Retrieval Query Language (SRQL) specifically designed for the enterprise KG, which is significantly different from SPARQL.

\textit{KGLac} \cite{Helal2021} follows a similar approach in that it is an external service that generates a central KG based on the extracted profiles to perform all data management. The data profiling is assisted by a machine learning model to represent each data source in an embedding space to perform a similarity search between tables or columns without revealing the raw data. 
Again, in the KG vertices represent data sources, i.e., tables, or columns, while edges represent relationships between these nodes. 
In contrast to \textit{Aurum}, KGLac relies on a custom extension of the RDF standard.
A specific ontology is used for querying with SPARQL to allow navigation and maintenance of the central KG. 
The system does not have a user interface, but relies on Python APIs to query a RDF database with direct integration into common data science pipelines.

\textit{CoreKG} \cite{Beheshti2018} is an open-source service developed for data lakes to curate stored data sources via REST APIs, which is composed of the extraction of metadata, their enrichment, linking, and annotation. 
Linkage in CoreKG means establishing a connection to an external KG such as WikiData as well as to other data sources. 
It is not a complete data lake, because it does not provide any storage or processing capabilities,
but its functionality can be easily added to existing data lakes using the REST API.

\textit{BARENTS} \cite{Stach2021} introduces an ontology-based method that enables analysts to model how data sets have to be pre-processed to transform them into meaningful knowledge. 
They introduce an ontology, not for data discovery but explicitly for pre-processing. 
Domain experts can use this ontology to describe which data transformations have to be applied to the raw data from source systems in order to make it exploitable in their use cases.  

\subsection{Commercial systems}
\label{sec:commercial-systems}
There exists a series of proprietary software stacks worth being mentioned, but due to the `paywall', the functionality and details are not fully revealed. 

The metadata model used by Bosch as described in the previous section is implemented using the \textit{Enterprise Knowledge Graph Plattform} by Stardog \footnote{\url{https://www.stardog.com/}} for storing and processing the semantic layer as a KG.
The \textit{Poolparty Semantic Suite}\footnote{\url{https://www.poolparty.biz/}} from the Semantic Web Company offers a wide range of service, accessible through (REST) APIs that can be exploited in conjunction with a data lake. 
The \textit{Anzo Semantic Data Lake}\footnote{\url{https://cambridgesemantics.com/solutions-3/asdl-2/}} offered by Cambridge Semantics focuses on adding context by first building a catalog of KGs from existing data sources in a data lake and then aligning those to semantic models based on business meaning. 

\begin{table*}
	\centering
	\caption{Comparison of data lakes with a semantic layer.}
	\label{table-datalakes}
	\begin{tabular}{|l||c|c|c|c|c|c|c|c|}
		\hline
		 &  \textbf{SL} & \textbf{SR} & \textbf{C/E} & \textbf{MI} & \textbf{IAL} & \textbf{TA} & \textbf{C} & \textbf{O} \\\hline
		 	
		\textit{Goods}\cite{DBLP:conf/sigmod/HalevyKNOPRW16} & \checkmark & \checkmark  & \text{\sffamily X}  & \text{\sffamily X} & \checkmark & \text{\sffamily X} & \text{\sffamily X} &  \text{\sffamily X} \\\hline
		
		\textit{Constance}\cite{constance} & \checkmark  & \text{\sffamily X} &  \text{\sffamily X} & \text{\sffamily X}  & \text{\sffamily X} & \text{\sffamily X} & \text{\sffamily X} & \text{\sffamily X}  \\\hline
		
		\textit{Aurum}\cite{Fernandez2018a} & \checkmark & \checkmark & \checkmark & \text{\sffamily X} & \checkmark & \text{\sffamily X} & \text{\sffamily X} & \checkmark \\\hline
		
		\textit{KGLac}\cite{Helal2021} & \text{\sffamily X} & \checkmark & \checkmark & \text{\sffamily X} & \checkmark & \checkmark & \checkmark & \text{\sffamily X}  \\\hline
		
		\textit{CoreKG}\cite{Beheshti2018} & \checkmark & \checkmark & \checkmark  & \checkmark & \checkmark & \checkmark & \checkmark & \checkmark \\\hline	
		
		\textit{BARENTS}\cite{Stach2021} & \text{\sffamily X} & \checkmark & \checkmark & \checkmark & \text{\sffamily X} & \checkmark & \checkmark  & \text{\sffamily X} \\\hline	
		
		\textit{Enterprise Knowldege Graph Plattform} by Stardog at Bosch & \checkmark & \checkmark & \checkmark  & \checkmark & \checkmark & \checkmark & \checkmark & \text{\sffamily X} \\\hline	
		
		\textit{Semantic Integrator} from the Semantic Web Company & ? & \checkmark & ? & ? & \checkmark & \checkmark & ? & \text{\sffamily X} \\ \hline
		
		\textit{Anzo Semantic data lake} & ? & \checkmark & ? & ? & \checkmark & \checkmark & ? & \text{\sffamily X} \\ \hline
		
		\textit{Semantic Layer} by AtScale & \text{\sffamily X} & \checkmark & \text{\sffamily X} & \checkmark & \checkmark & \checkmark & \text{\sffamily X} & \text{\sffamily X}\\ \hline
		
		\textit{Semantic Layer} Dremio LakeHouse Plattform & \text{\sffamily X} & \checkmark & \text{\sffamily X} & \checkmark & \checkmark & \checkmark & \text{\sffamily X} & \text{\sffamily X} \\ \hline
	\end{tabular}
\end{table*}

The \textit{Semantic Layer} by AtScale\footnote{\url{https://www.atscale.com/}} follows a different approach in that it does not apply Semantic Web technologies at all. 
Instead, it focuses on data integration through data modeling in a canvas, where users can establish relationships and hierarchies between heterogeneous data sources from different storages without ever replicating or moving any data, but only via reference. 
The \textit{Semantic Layer} complements cloud data platforms (such as DataBrick's \textit{LakeHouse} \cite{Zaharia2021}) and provides a platform for integration and orchestration of the platform's query and transformation engines.
This bridges the gap from the data sources to business intelligence tools and AI applications. 
A typical workflow involves selecting data sources from cloud storages or a data lake, modeling their schemas, establishing relationships, and exporting the model to an analytics platform (such as \textit{Power BI} or \textit{Tableau}). 
All the connection details to the different storage platforms are hidden by abstraction and users may start generating insights based on the semantic model created by them.

A similar approach is followed by the \textit{Dremio Lakehouse Platform}\footnote{\url{https://www.dremio.com/resources/demos/dremio-semantic-layer/}}. Just like the \textit{Semantic Layer} by AtScale, \textit{Dremio}'s semantic layer manifests into a hierarchical structure that exposes business representations of an organization's data assets to enable exploration.

\subsection{Comparison \& discussion} 
\label{chapter-compare-datalakes}

We identify criteria to compare and to check the comprehensiveness of the existing systems summarized in \Cref{table-datalakes}. The table is limited to concrete data lake systems and does not list approaches that just propose a metadata model.
\begin{enumerate}[leftmargin=0.5cm]
    \item \textbf{SL:} \textit{Semantic labels} denotes the possibility to connect schema entries with instances in an external KG, either a domain KG to be uploaded or an online KG such as \href{https://schema.org/}{schema.org}. 
    
    \item \textbf{SR:} \textit{Semantic relationships}, i.e., any form of hierarchical, generic, or pre-defined semantic relationships (semantic connections between data sets, e.g., for provenance or governance). 
    It should be noted that none of the existing data lake systems provides advanced modeling features such as those depicted in \Cref{fig-intro-overview}.
    
    \item \textbf{C/E:} \textit{Central KG graphs vs external KGs}. Here we distinguish whether the underlying metamodel itself is modeled directly as a KG (i.e., relationships are limited to those defined in this KG). A more flexible system would allow an external KG, and any concepts and relationships from this KG can be used in the semantic model. A checked field indicates that the metadata model uses a central, predefined KG.  
    
    \item \textbf{MI:} \textit{Metadata interoperability} 
    This denotes whether the authors have considered the possibility of importing or exporting metadata. i.e., serialization of the semantic model.
    This is strongly related to criteria \textbf{C} since serialized models  are ported to different systems more easily.
    
    \item \textbf{IAL:} \emph{Initial automatic semantic labeling} denotes the functionality to generate semantic labels automatically from the schema at the time of ingestion.
    
    \item \textbf{TA:} \textit{Technical abstraction}. This denotes whether the authors have explicitly considered the technical abstraction of the semantic modeling as a requirement to increase usability for unfamiliar users. This means that technical details are hidden from the users, e.g., a user does not need to be familiar with RDF.
    
    \item \textbf{C:} \textit{Compatibility with Semantic Web technology}. This criterion checks whether semantic labels are modeled in a language of the Semantic Web (e.g., RDF, OWL). 
    
    \item \textbf{O:} \textit{Open Source}. This criterion checks whether there is an open-source implementation available.
\end{enumerate}

%% file: chapters/SM.tex
\section{Semantic Modeling for Data Integration}\label{chapter-automatic}
Semantic modeling targets at providing a semantic description of data sources and is not restricted to data lakes, but can be applied to data management in general. Nevertheless methods for enriching metadata are very relevant in this context, because they increase the usability of a data lake.
The semantic model acts as a bridge between the data source and a semantically rich KG,
which can be applied in data integration or ontology-based data access. 
In this section, we first focus on approaches for creating semantic models semi-automatically, potentially supported by human input in a graphical user interface (GUI). 
Manual definition of semantic models is possible, but is not feasible for a large number of heterogeneous data sets as expected in data lakes. 
Thus, automating the generation of semantic labels and models helps to reduce manual efforts in a semantic data lake system \cite{DBLP:conf/jowo/BraunCF19a}. 
In the following sections, we investigate the current state of such proposals and conclude with a comparison and discussion.

\subsection{Semantic modeling systems} 
The systems mentioned in the previous section focus on \emph{semantic labeling}, i.e., enable the annotation of data sets with elements from internal and external ontologies. 
\emph{Semantic modeling} is the next step in semantic enrichment: the associated ontology elements are enriched with further semantics to reflect the relationships represented in the data sets (see \Cref{fig-intro-overview}).  

\textit{ESKAPE} \cite{Pomp2017} is a semantic data platform handling the full data management process from ingestion to extraction supporting heterogeneous data sources. It introduces an intuitive interface where users can create semantic models. 
In \cite{pomp2018applying}, the authors report the integration of \textit{ESKAPE} into a data lake system in a production setting. 
Instead of just dumping raw data into the data lake directly, they use \textit{ESKAPE} during data ingestion to enable the semantic annotation of data sets to perform semantic data integration.
To combine the defined semantic models, \textit{ESKAPE} also allows the integration of external KGs, such as WordNet, BabelNet, or domain-specific ontologies.
Elements of the KGs can be reused during the modeling process and therefore serve as anchors when relating different distinct semantic models and data sets. 
 
\textit{Karma} \cite{Gupta2012} is a data integration tool that enables users to ingest data as well as as a vocabulary defined as an OWL ontology to create a corresponding semantic model. 
\textit{Karma} supports multiple data formats and proposes an initial semantic model which the users can edit using a GUI. 
\textit{Karma} is an important pioneering work being one of the first to introduce key technologies such as conditional random fields (CRF) and an algorithm to solve a Steiner tree problem 
for learning of semantic labels and to discover relationships among the schema elements of a source.
Once the semantic modeling based on the ontology is completed, Karma integrates data sets based on the defined semantic model. 
The resulting semantic models can be exported as a RDF file.
In contrast to \textit{ESKAPE}, \textit{Karma} provides initial semantic labeling suggestions, but it also requires users to have in-depth knowledge about technical details, e.g., the RDF format. Furthermore, it focuses on automatically creating semantic models based on a fixed ontology per use case where the ontology does not evolve over time.
\textit{ESKAPE} claims to be easier to use because it hides the technical details; however, there is no implementation available to compare the usability. 

Recently, the open-source platform \textit{PLASMA} \cite{Paulus2021a} has emerged from \textit{ESKAPE} as well as several other notable works, described in more detail in the next section. 
Paulus et al. introduce a phase called \textit{semantic refinement} (see \Cref{fig-DL-PLASMA}) in which the human is responsible to improve the quality of the semantic model by manually checking, validating, correcting, selecting, or exchanging concepts.
\textit{PLASMA} comes with a semantic concept recommendation framework to suggest semantic concepts extracted from public KGs as well as any proprietary source (e.g., a company's KG). 
The semantic models can be created, deleted, searched, and exported as RDF in Turtle syntax.
Central to the whole framework is the idea of continuously evolving ontologies.
As data sources need to be added or updated, the created semantic model must be extended or adapted to reflect these changes.
To address this issue, Pomp et al.~\cite{pomp2019enabling} propose an approach featuring an evolutionary KG that consists of an internal growing universal KG and data source specific mappings.
A user, who creates a semantic label, is enabled to define and use concepts and relationships in their semantic models that were previously unknown to the KG.
This is, for instance, the case if the concept the user wants to use is missing in the current conceptualizations. 
This novel knowledge must be integrated into the ontology of the universal KG. 
For this, Pomp et al.\ developed a strategy for the controlled and evolutionary development of the KG which evaluates the relationships and concepts that a user selects or introduces, and validates if the user introduces contradictions to the current KG. 
The framework identifies additional relationships between concepts that improve the density of the underlying ontology and contains validation logic to eliminate inconsistencies.

The dataspace management system On-demand Data INtegration (\textit{ODIN}) \cite{Nadal2019} allows data integration by virtually querying heterogeneous data sources grounded on KGs. 
\textit{ODIN} automatically extracts the schemata from (semi-)structured data sources, translates them into a canonical data model, aligns their schemata, and generates target-specific metadata from them. 
In the spirit of semantic refinement, it relies on a supported user feedback process for incrementally merging the semantics into a dataspace-wide ontology\footnote{A demonstration of the system is available at \url{https://www.essi.upc.edu/~snadal/odin.html}}.
\textit{ODIN} supports an advanced visual query mechanism by relying on \textit{WebVOWl}\footnote{\url{http://vowl.visualdataweb.org/webvowl.html}}, which is an open-source application for  interactive visualization of ontologies. This ontology-mediated query interface allows selecting nodes of interest from any available semantic model by marking them graphically in order to generate a SPARQL query.
Data sources associated with the marked nodes represent the connection point to the data sets; wrappers encode the query to extract their data and expose a first-normal form relation of their schemata.


\subsection{Automated assignment of semantic labels}
The technique of bootstrapping an initial ontology from (relational) databases and directly establishing a mapping between the databases and the created concepts is present in several systems. 
For example, \textit{Optique} as well as its successor \textit{OnTop} (see  \cref{chapter-OBDA-SQL}) provide such a capabilities. 
An early benchmark for evaluating the quality of the generated mappings in the context of ontology-based data access (OBDA, see \cref{chapter-OBDA}) is presented by \textit{RODI} \cite{Pinkel2018}. 
Influential works include the \textit{SemanticTyper} presented by Ramnandan et al. \cite{Ramnandan2015}, who apply heuristic rules (a TF-IDF-based approach for textual data and the Kolmogorov-Smirnov statistical hypothesis test for numeric data) 
as well as the Domain-Independent Semantic Labeler (\textit{DSL}) by Pham et al. \cite{Pham2016}, which presents a machine learning approach around similarity measures.

As mentioned previously, interested readers are referred to the survey by Liu et. al. \cite{liu2022tabular} for a thorough and comprehensive discussion of proposals for STD's up until the year 2021 including the numerous developments based on deep learning. Here, we want to include one more recent approach called \textit{Tab2KG} (\cite{Gottschalk2022}, 2022), which follows an entirely different approach in the sense that it can interpret previously unseen data on the basis of arbitrary domain KGs. 
Interestingly, \textit{Tab2KG} uses RML to serialize semantic labels possibly useful to perform OBDA (see \cref{chapter-OBDA})
They introduce so-called semantic profiles for domain ontologies and data tables to facilitate effective semantic table interpretation.
For creating a domain profile, the domain KG, that contains representative values for the data type relations in the target domain, is transformed into a feature vector containing a set of statistics, computed using all literals coming from to set of data type relations. 
For creating a data table profile, a feature vector of descriptive values is computed which includes data types and basic statistics (completeness, mean, standard deviation, skewness, histogramms etc.).
The selection is motivated by the expected feature effectiveness for semantic table interpretation, i.e., matching the domain and data table profiles and can be extended to include relevant domains-specific characteristics.
The profiles are generated automatically and described using the DCAT2 and the SEAS3 vocabularies to enable their reusability.
Then, Tab2KG uses the domain and data table profiles to perform the semantic type detection using a novel one-shot learning approach.
In contrast to domain-specific approaches such as \textit{DoDuo}\cite{DBLP:conf/sigmod/SuharaL0ZDCT22, liu2022tabular}, \textit{Tab2KG} uses a Siamese network 
that generalizes towards unseen data type relations by inducing a metric that represents the domain-independent similarity between two input feature vectors (e.g., between an unknown and a known sample). 
The similarity between a column and a data type relation is predicted based on the experience of the similarity of other profiles learned earlier.
Given a set of candidate column mappings with their respective similarity scores for each column in the input data table, they map each table column to a data type relation in a greedy manner..

\subsection{Automated semantic model generation} \label{chapter-semantic-model-generation}
While the previous approaches focus on semantic labels solely, there also exists some that aim to construct entire semantic models (some including semantic label generation as well) and the field has attracted increasing interest lately.

Vu et al. \cite{Vu2019} present \textit{PGM-SM} which uses probabilistic graphical models to create semantic models. From previously generated semantic models and the detected semantic labels, a conditional random field is trained to distinguish between good and bad semantic models. The graph is then used to identify possible semantic models and the best k out of them are selected using a scoring function based on probabilistic graphical models.
In follow-up work, \cite{Vu2021} they recognize that despite being flexible in choosing a target ontology, the approach requires users to label enough data sources before the systems can achieve good performance and this issue is more profound the larger the target ontology is. For this reason, they shift the supervised problem towards utilizing large-scale KGs, such as DBPedia and WikiData.
For this, they present a novel probabilistic approach for automatically building semantic descriptions of Wikipedia tables leveraging hyperlinks and existing knowledge in Wikidata to construct a graph of possible relationships in the table and its context.
To assess the effectiveness of the method, it is evaluated on a data set from the SemTab2020 challenge as well as another data set with 250 Wikipedia tables with their semantic descriptions built using the Wikidata ontology.

Further notable works related to the SemTab2020 challenge are MantisTable \cite{Avogadro2021}, bbw \cite{Shigapov2020}, and MTab4WikiData \cite{Nguyen2020} each of them aiming at the automatic generation of semantic models for data from Wikipedia or Wikidata.

In the spirit of \textit{CoreKG}, Feng et al. \cite{Feng2021} present \textit{ASMaaS} (Automatic Semantic Modeling as a Service), which is an external service for data lake vendors and users to  generate a semantic model for their data sets. The paper describes a service-oriented architecture including annotating training data, training the machine learning models, and predicting an accurate semantic model for new data sources. 

Futia et al. \cite{Futia2020} present an approach called \textit{SeMi} (SEmantic Modeling machIne) based on graph neural networks (GNNs), that was inspired by the work of Taheryan et al. \cite{Taheriyan2013, Taheriyan2016, Taheriyan2016a}, Knoblock et al. \cite{Knoblock2012}  (who are also co-founders of \textit{Karma}) as well as the \textit{SemanticTyper} by Ramnandan covering the process of semantic modeling. 
They adopt the exploitation of linked data repositories as background knowledge similar to Taheriyan et al. 
They replace the manual extraction of compound features (e.g., complex graph patterns to represent semantic relationships of different lengths) by a graph neural network that automatically learns latent features for entities and properties, encoding them in a vector space exploiting the local neighborhood structures within the linked data graph. 
In their processing pipeline, the Semantic Model Builder generates an initial semantic model. The proposed Steiner tree within the graph includes the shortest path to connect semantic type classes. 
However, it does not necessarily express the correct semantic description of the target source. 
For this reason, a refinement process is required in order to identify a more accurate semantic model. 
The semantic model refinement (notice, that here an automatic process is applied as opposed to the one in \textit{PLASMA}, cf. \Cref{fig-DL-PLASMA}) requires preparing data sets as input of the deep learning model. 
The graph neural network's main goal is to reconstruct the linked data edges using the latent representation of entities and properties.

Burgdorf et al. \cite{Burgdorf2022} present \textit{VC-SLAM}, a corpus that allows the evaluation and comparison of semantic labeling and modeling approaches across different methodologies. 
It contains 101 data sets composed of labels, data, and metadata, as well as corresponding semantic labels and a semantic model. The semantic labels and the semantic model 
are manually refined by human experts using an ontology that was explicitly built for the corpus. 
While the first two directions in the semantic labeling process (see \cref{chapter-OBDM}), i.e., data- \& schema-driven, have been addressed by the majority of the works presented so far, the third direction has attracted only little attention so far. 
Metadata-based semantic labeling regards all available additional pieces of information on a data set that might contribute to the semantic labeling of the data. For example, structured data like CKAN standards and any additional metadata within a database, like comments or textual data documentations stored in documentation tools or Wikis could be used to enhance the semantic labeling process.
They have tried to evaluate \textit{SeMi} against their datasets, but were unable, due to the fact that they have created a novel domain ontology which is not reasonably big enough to train the GNN with background data.
For this reason, they include additional exploitable textual data documentation.

Burgdorf et al. present also \textit{DocSemMap} \cite{Burgdorf2022a} a novel approach that utilizes textual data documentations of data sets as an additional source for the creation of semantic mappings.
It utilizes Natural Language Processing (NLP) techniques jointly with structured data to perform semantic modeling. 
They show that especially when compared to data-driven methods, there are different cases where \textit{DocSemMap} can map concepts that would not otherwise be identified.
Hence there is great potential for performing semantic mapping based on textual data documentations.

In \cite{Ramirez2021}, Ramirez et al. present Relational Natural Language Inference (\textit{RLNI}) aiming to provide explainable data exploration on data lakes. 
They address the fact that many frameworks tend to build on similarity metrics that stop short of providing a clear explanation as to how an identified data set relates to a provided target. 
In order to tackle this problem they focus on based on Natural Language Inference (NLI), which has a limited set of relationships (i.e., equivalence, forward entailment, reverse entailment, negation, alternation, cover, and independence), and provide an unsupervised analysis as well as a supervised alternative, that requires training data both based on pre-trained language models. 
The approach lacks to incorporate the semantic expressivity inherent to full-fledged ontologies, but rather provides semantics and explainability to the problem of exploration in data lakes which is recognized as insufficient in comparable works such as \textit{Aurum} \cite{Fernandez2018a, Fernandez2018}.

Haller et al. \cite{Haller2019, Haller2020} study the whole problem from a different angle. In particular, they analyze SQL queries written by data analysts, who already understand the semantic relationships of the heterogeneous data sources. Their prototype called \textit{Pharos} connects to an existing SQL session of any program that has a JDBC interface, and records and analyzes all queries in the background. It extracts knowledge fragments from SQL queries and represents them in an RDF-based KG for which they have designed a new ontology. Examples include foreign/primary key constructs extracted from JOIN queries or log entries of the form ``attribute salary has been cast to the data type long'' from which it can be concluded that the salary column can be summed. 

\subsection{Comparison \& discussion}
The comparison is divided into two parts. We first review complete modeling platforms, that include methods for semantic labeling and/or modeling, as well as a graphical user interface to allow the user the adaptation of the created artefacts.  
For this category, we have identified several criteria:
\begin{enumerate}[leftmargin=0.5cm]    
    \item \textbf{IAM:} \textit{Initial automatic semantic model}. In contrast IAL in \cref{chapter-compare-datalakes}, this denotes here the functionality to generate entire semantic models automatically from the schema after the ingestion of a data source. 
    This includes the first three phases in \Cref{fig-DL-PLASMA}. 
        

    \item \textbf{EX:} \textit{Support for external KGs}. Is the approach capable of handling arbitrary KGs or fixed to a predefined KG?
    
    
    \item \textbf{M:} \textit{Maturity}. Here we check whether the system has been reported to be successfully utilized in a Big Data project. 

    
    
    \item \textbf{C:} \textit{Compatibility with Semantic Web technology}. This criterion checks whether semantic labels are modeled in a language of the Semantic Web (i.e., RDF or OWL). 
    
    \item \textbf{O:} \textit{Open-source software} is available.
    
\end{enumerate}

The comparison is summarized in \Cref{table-semanticmodellingplattforms}.

\begin{table}[tb]
	\centering
	\caption{Comparison of semantic modeling platforms.}
	\label{table-semanticmodellingplattforms}
	
	\begin{tabular}{|l||c|c|c|c|c|c|c|} \hline
		
		 & \textbf{IAM} & \textbf{SR}  & \textbf{EX}  & \textbf{M}  & \textbf{C} & \textbf{O}  \\ \hline
        
        \textit{ESKAPE} \cite{pomp2018applying} & \checkmark & \text{\sffamily X} & \checkmark  & \checkmark   &  \text{\sffamily X} & \text{\sffamily X}  \\ \hline
        
        \textit{KARMA} \cite{Szekely2015} & \text{\sffamily X} & \checkmark & \checkmark & \checkmark   &  \checkmark & \checkmark \\ \hline
        
        \textit{PLASMA} \cite{10.1145/3459637.3481995} & \checkmark & \checkmark & \checkmark & \checkmark  & \checkmark  & \checkmark \\ \hline
        
		\textit{ODIN} \cite{Nadal2019} & \checkmark & \checkmark & \checkmark & \text{\sffamily X} & \checkmark  & \checkmark  \\ \hline
		
	\end{tabular}
\end{table}

It should be noted that all of the mentioned approaches focus on relational/tabular or at least
(semi-)structured data (including JSON and XML) only. The heterogeneity inherent to Big Data applications is still far from being covered for automatic semantic modeling approaches. 
In terms of GUI-based manual modeling \textit{PLASMA} displays a mature solution, proven to be applicable in production with many built-in features which can be tested publicly.\footnote{\url{http://plasma.uni-wuppertal.de/modelings}}

Semantic Type Detection has gained in research momentum lately, but the field is far from a truly mature solution. 
The latest solutions based on deep learning achieve impressive accuracies when evaluated on their specific test data sets, but considering the fact that any labeled data set to performed supervised learning (such as \textit{WikiTables}\footnote{\url{http://websail-fe.cs.northwestern.edu/TabEL/}} or \textit{GitTables} \footnote{\url{https://gittables.github.io/}}) can only contain a limited amount of semantic types, its usefulness in practise becomes very limited.
They utlimately suffer from a closed-world assumption as in the case of \textit{DoDuo} \cite{DBLP:conf/sigmod/SuharaL0ZDCT22}: their final model can distinguish only between 255 different types. 
In order to sensitize the model for a different domain ontology (other than WikiData), it is required to train the model on a manually labeled data set containing corresponding semantic types.
Furthermore, because they use a language model, the detection performs poorly on numeric data so far and most approaches are limited in dealing with high-dimensional and large data sets. 
Taking up \textit{DoDuo} again, it has been trained on tables with an explicit maximum of 10 columns and 32 tokens in sequence length.
In contrast, the embedding approach in \textit{DAGOBAH} or the heuristic strategies by \textit{MTab} and \textit{JenTab} \cite{DBLP:conf/semweb/ChabotLLT19, Nguyen2019,Abdelmageed2021} may possibly predict arbitrary semantic types, because here semantic types are extracted via look-ups, that are not limited to specific targets. 
However, both the strategies evaluate their algorithms on excerpts of cross-domain KGs such as WikiData and DBpedia and are not specifically designed to work with a specialized conceptualization, such as a domain ontology. This is particularly important for the applicability to data lakes, as they are often utilized in a specific domain (e.g., smart manufacturing \cite{Dibowski2020a, Mami2020} or the energy sector \cite{gagliardelli2022ecdp}) and specialized knowledge is needed for in-depth data understanding. 

In terms of automated modeling, \textit{SeMi}'s approach relying on KG embeddings computed with GNNs is an advanced solution, but its semantic type detection, the first and most crucial step in the pipeline, is not generic but tailored towards the specific data sets \cite{Futia2020}.


%% file: chapters/obda.tex
\section{Ontology-based Data Access}
\label{chapter-OBDA}

With semantic labels and semantic models, a semantic description of the data sources is available; it is hence reasonable to use these semantics for data integration and querying. 
This is the topic of \emph{ontology-based data access} (OBDA) which has been developed since the mid-2000s \cite{DBLP:journals/jods/PoggiLCGLR08} with the aim of facilitating access to various types of data sources, using an ontology as the common data model. 
The setup can be described as a set of existing data sources forming the data layer, and the goal is to build a service on top of this, aiming at presenting a conceptual view of data to the clients. 
The ontology describes the domain of interest and is usually developed independently from the data layer. 

A challenge in this scenario is the combination of different abstraction levels and formalisms. Whereas ontologies model the conceptual level, use open-world assumption and are based on description logics, the data sources are described by schemas on the logical level which focus on the structure of the data and use the close-world assumption. 
By posing the query to the ontology, the primary advantage of OBDA is that users can query various data sources without the need to know how the data is organized and where it is stored. 

Fathy et al.\ \cite{fathy2019unified} identify two main approaches for applying OBDA: 
(1) \textbf{materialization}, meaning all source data sets are converted into a common data format (e.g., RDF or relational tables) and stored in one data repository, or
(2) \textbf{on-demand translation} (or rewriting) of queries (ususally expressed in SPARQL) into the required query language of the data sources.
In this survey, we emphasize the applicability to Big Data; hence, scalability is a strong requirement. Materialization approaches can be challenging and expensive when dealing with large data sets.
Hence, only query translation approaches are feasible in data lakes, albeit the fact that it is significantly more complex, because it requires profound theoretical consideration.
It is part of the larger domain of data federation \cite{gusystematic} which addresses the problem of uniformly accessing multiple, possibly heterogeneous data sources, by mapping them into a unified schema. 
The common data model used for querying can be also expressed in other schema languages. 
Earlier approaches for data integration used the relational model and translated the data sources into a common relational schema \cite{DBLP:conf/pods/Lenzerini02}. 
Query rewriting in this relational setting has been well studied; we consider the more complex scenario of query translation with ontologies over heterogeneous sources.

\begin{figure}
    \centering
    \includegraphics[width=1\columnwidth]{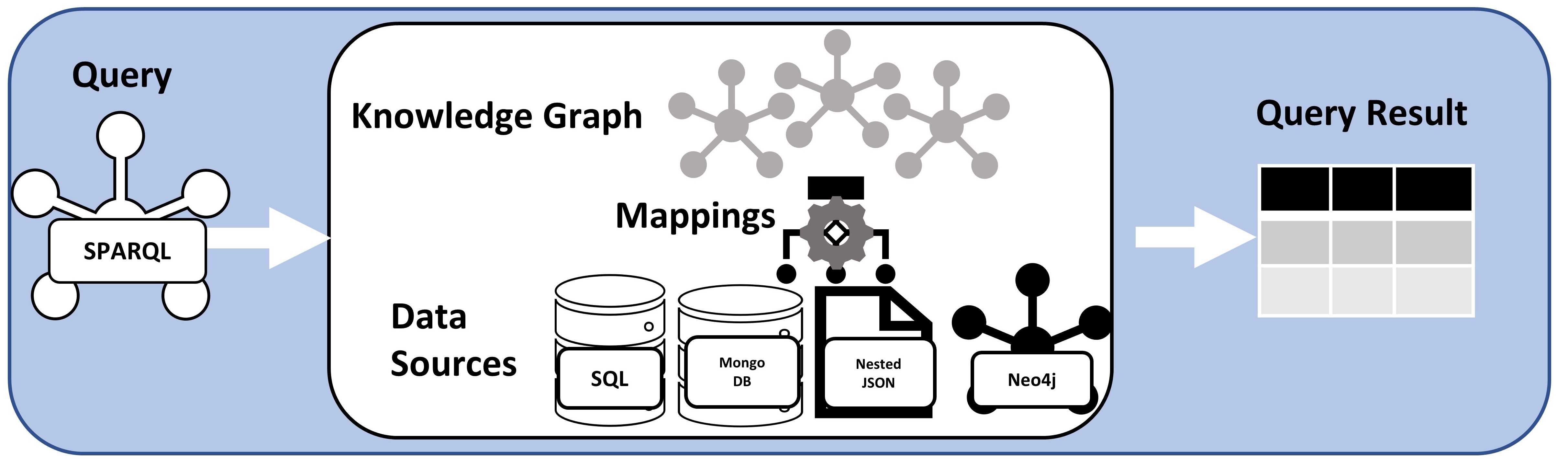}
    \caption{Illustration of Ontology-Based Data Access}
    \label{fig-OBDA}
\end{figure}

Query translation \cite{Mami2019a} is hence strongly related and essential for any effective OBDA framework to deal with the variety of heterogeneous data sources in a data lake. 
In OBDA, on-demand query translation exposes the content of data sources as RDF triples, using classes and predicates from a KG. 
The RDF triples are not materialized, they are part of a virtual knowledge graph (VKG), which means that data remains in the data sources instead of being stored in some common database.
A VKG system has the following components (see \Cref{fig-OBDA}): (a) queries that describe user information needs, (b) an ontology with classes and properties, (c) mappings to the data sources, and (d) a collection of data sources. 
The W3C published recommendations for languages for components (a)–(c): SPARQL, OWL 2, and R2RML\footnote{\url{https://www.w3.org/TR/r2rml/}}, respectively.

Mapping creation and management is probably the most complicated OBDA design-time task, as the mapping specifies the semantics of the data sources in terms of the ontology and bridges the typically large conceptual gap between the source schema and the ontology.
By applying knowledge representation and automated reasoning techniques, an OBDA system uses ontologies and mappings to reformulate the SPARQL queries into source-specific queries. 
A formal framework is presented by \cite{DBLP:conf/ijcai/XiaoCKLPRZ18}, including query answering, mapping management and analysis, and extensions of the classical OBDA framework. 
Here, we omit the theoretical details of query rewriting and focus on the practical application for data integration in the Big Data regime. 

A comprehensive survey on query translation methods by Mami et al.\ is given in \cite{Mami2019a}, including a classification, a map for different language translation paths, and a historical timeline. 
While computer scientists have been looking for the holy grail of data representation for decades (logic, ER/relational, XML, graph, NoSQL), their work reinforces the view that no single data representation scheme fits to all use cases.
Mami et al.\ further identify several interesting shortcomings: 
First, they recognize that many translation methods fail to support the more sophisticated operations such as different join types and temporal functions in SQL, or blank nodes, grouping, and binding in SPARQL.
Then, they state that there a many well-founded and defined query translation frameworks, from the query translation process to the various optimization strategies; however, they would hardly represent real-world queries, but mostly simple queries. 
Use-case-driven translation methods would be more helpful to reveal useful query patterns and to evaluate the translation methods and optimizations on real-world data.
They state that there is a wide variety in the evaluation frameworks used by each of the query translation methods and a need for a unique standardized benchmark specialized in evaluating and assessing query translation aspects.
In their final remarks, they discuss candidates for a `universal' query language. 
Based on their findings exploring various query translation methods, they see SQL and SPARQL as the most suitable languages to act as a `universal' language for realizing heterogeneous data integration. 

\subsection{OBDA over relational data}\label{chapter-OBDA-SQL}

Some of the earliest systems performing OBDA include \textit{Mastro} (2011, \cite{Calvanese2011}),  \textit{UltraWrap} (2013, \cite{Sequeda2013}) and \textit{MorphDB} (2014, \cite{Priyatna2014}). \textit{Optique} (2013 \cite{Kharlamov2013}) was the first successful OBDA system designed for Big Data scenarios. 
Its underlying architecture mainly uses \textit{Ontop} (see below) to manage data access using R2RML mappings, and Exareme \cite{Chronis2016}, a research prototype which acts as a back-end query execution component handling large-scale data processing tasks. 
A major drawback of these early systems is the lack of horizontal scalability.

The \textit{Ontop} framework for OBDA \cite{Bagosi2014} is likely to be one of the most mature open-source\footnote{https://github.com/ontop/ontop} OBDA systems, which has come a long way since its beginnings in 2009 \cite{Bagosi2014, DBLP:conf/dlog/0001LKKKDCCCB20}. 
It is the result of an active research and development community and has been adopted in many academic and industrial projects \cite{Kharlamov2017,Kharlamov2017a, Mansfield2021}. 
Because \textit{Ontop} moved towards supporting the W3C recommendations for SPARQL and R2RML, new challenges emerged, which required the development of a core data structure called intermediate query (IQ), an algebra-based data structure that unifies both SPARQL and relational algebra. 
IQ is a uniform representation both for SPARQL queries and for SQL queries from the mapping. 
When the query transformation (rewriting and unfolding) is complete, the IQ expression is converted into SQL, and executed by the underlying relational database management system (DBMS). 
\textit{OnTop} aims to shift almost all query processing to the DBMS and performs only the topmost projection, which typically transforms database values into RDF terms. 
Sophisticated optimization techniques such as redundant join elimination, using primary and foreign keys, efficient duplicate elimination \cite{DBLP:conf/dlog/BilidasK18}, and pushing down joins to the data-level \cite{Rohde2020} as well as exploiting integrity constraints to enhance query completeness and performance \cite{ChavesFraga2021} have been extensively studied during the past decade to make sure that queries can be efficiently processed by the underlying DBMS.

With a recent extension called \textit{Ontop4theWeb} \cite{Bereta2021}, the framework can now be evaluated on-the-fly against web data using SPARQL without materializing it into a DBMS first. 
\textit{Ontop4theWeb} implements an approach for posing SPARQL queries on top of non-RDF Web data on-the-fly. 
To achieve this, virtual table operators are embedded in the SQL queries that are included in R2RML mappings. 
They specify which part and source of Web data will be fetched and how they will be mapped to virtual RDF terms. 
The specification these mappings creates the usual overhead, but they need to be specified only once unless the schema changes.
Although no explicit materialization of data is performed and the responses of web-native and NoSQL JSON APIs can be obtained, \textit{Ontop4theWeb} still relies on the relational data model and relies on a relational database system in the backend.

\textit{Ontopic Studio}\footnote{\url{https://ontopic.ai/}} is a commercial spin-off of \textit{Ontop} that provides a GUI to ease the creation and search of mappings via a mapping editor.
The \textit{Ontop} framework has recently attracted interest for an application in data lakes:
Schwade et al.\ \cite{Schwade2020} show a successful integration into a data lake by adding a semantic layer based on the \textit{Ontop} framework. 

The \textit{Chimera Suite} \cite{Belcao2021} extends the native \textit{OnTop} systems by two components: 
(1) Ontop$_{Spark}$, an extension that enables \textit{OnTop} to perform OBDA on relational data using Apache Spark effectively translating \textit{SPARQL} queries into \textit{SparkSQL} and 
(2) a JDBC endpoint to connect \textit{Spark Thrift Server} with \textit{OnTop} that can direct the \textit{SparkSQL} queries to an HDFS filesystem.
Belcao et al.\ point out the current gap between Big Data technologies and the semantic world.
On the one hand, knowledge engineers work with OWL and RDF to create knowledge graphs. On the other
hand, data engineers define ETL processes to provide the source data in form which can be consumed by data scientists.
Then, data scientists need to combine these two perspectives to provide interpretable analytics results.
To tackle this problem, Belcao et al.\ introduce a comprehensive pipeline, where data scientists write SPARQL queries and send them to Jena Fuseki in a Jupyter Notebook using the PySPARQL library to get a semantically enriched response. 
Using the mappings and the ontology, the query is re-written into SQL. 
Ontop$_{Spark}$ accesses the data stored as Parquet files in HDFS using Apache Spark. 
The SPARQL response returned from Ontop$_{Spark}$ is further enriched by Jena Fuseki using the knowledge graph, and the final results are sent back to the Jupyter notebook.

Belcao et al.\ further describe in \cite{Belcao2021} a real-world deployment at Italy's largest public company for industrial research in the energy sector \cite{bionda2019smart}. 
In this scenario, they developed a KG containing around 7 million triples representing Milan’s metropolitan area storing the network topology and all the information about the various pieces of electrical equipment.
The \textit{Chimera Suite} enables the data science team to make analytical predictions by querying both the KG and the Big Data repository. The knowledge engineers can access the graph and retrieve the results of the data scientists’ analysis.

\textit{Ontario} \cite{Endris2019} is a query processing engine over heterogeneous data sources in a data lake utilizing VKGs. \textit{Ontario} translates queries from a global querying mechanism, e.g., SPARQL, to the underlying native query mechanism of the data sources. 
Therefore, the system  transforms raw data to RDF on demand by applying mapping rules based on R2RML. \textit{Ontario} utilizes custom extensions of the RDF and SPARQL standards, where compliance with W3C standards is unclear.

\textit{Obi-Wan} \cite{Buron2020} is built on top of the \textit{Tatooine} \cite{Bonaque2016} mediator for heterogeneous sources and \textit{Graal}, a toolkit for query answering in knowledge bases  \cite{Baget2015}.
The approach relies on mappings that are Global-Local-As-View (GLAV) as opposed to Global-As-View (GAV), which are used, for example, in \textit{Ontop}, \textit{MorphDB} and \textit{Mastro}.
GLAV mappings are considered to be more flexible in scenarios with evolving data sources \cite{DBLP:conf/pods/Lenzerini02}. In GAV mappings, a change in a data source might require a change in the global schema and all source mappings. With GLAV mappings, the sources are described independently of each other with respect to the global schema.

This was also recognized by the work of Nadal et al.\ \cite{Nadal2023} who propose a framework for data integration entirely based on graphs utilizing GLAV mappings. 
A prototype of the rewriting algorithms is incorporated into \textit{ODIN} (see \cref{chapter-automatic}). 
The proposed query language is based on coverings of a graph representing the global schema and does not require users to define join conditions. The query is visually represented as graph. Required join operations on the data are generated by the rewriting algorithm.
Encoding all required metadata (i.e., global schema, source descriptions, mappings, and queries) as graphs simplifies the interoperability among them and allows rewriting algorithms to efficiently identify the relevant sources.

\textit{PolyWeb} \cite{Khan2019} relies on R2RML and its proposed successor RML\footnote{\url{https://rml.io/specs/rml/}}. 
\textit{PolyWeb} performs a predicate-based source selection where a set of relevant data sources on the Web (RDF \& CSV) for a given query are discovered by matching predicates used in a basic graph pattern and the input data sources. 
A SPARQL query provided as input is translated into a native query of the underlying system (CSV files, relational databases, RDF triple stores are considered) and executed on this system.

\subsection{OBDA over non-relational data}\label{chapter-OBDA-NoSQL}
There are four main groups of NoSQL Database Management Systems (DBMS) \cite{fathy2019unified}: document-oriented, column-oriented, graph-oriented, and key-value stores. 
Each of these DBMSs rely on different data models and query engines, which complicates the task of creating a uniform OBDA system in this case.
Because of increasing popularity and the lack of unified query languages and data models for these systems, there is growing interest in applying OBDA in this context. 
Systems based on VKGs promise to provide the desired unified semantic view of data sets.
Early papers include \cite{Cure2011, Cure2013} and different languages have been proposed to extend  the R2RML mapping language. 

RML \cite{DBLP:conf/www/DimouSCVMW14} is such a proposed extension to deal with heterogeneous data sources. 
There are proposals to extend RML with functions and operations described uniformly, unambiguously, and independently of the technology used for implementation.
Examples for those include the Function Ontology (FnO) \cite{Meester2017} and FunUL \cite{Junior2016}. 
While RML extends R2RML to schema transformations, the combination of RML with FnO extends R2RML with respect to data transformations.
RML is widely spread and implemented in existing systems, and many proposals exists that aim to ease its use. 
YARRRML \cite{Heyvaert2018} simplifies the use of both R2RML and RML by a human-readable text-based representation for mapping rules expressed in YAML\footnote{\url{https://yaml.org/spec/}}, a widely used human-friendly data serialization language.
The RML Editor \cite{Heyvaert2016}, RMLx \cite{aryan2017rmlx} Map-On (only for R2RML \cite{Sicilia2017}) and MapVOWL \cite{Heyvaert2018a} provide visual support to help users in generating Linked Data from raw data by defining RML mappings.
In \cite{DBLP:journals/peerj-cs/Garcia-Gonzalez20} the authors conduct a usability experiment on three different languages: ShExML \cite{DBLP:journals/peerj-cs/Garcia-Gonzalez20}, YARRRML and SPARQL-Generate \cite{DBLP:conf/ekaw/LefrancoisZB16}. 
RML was left out, because of too high verbosity.
Their prototype translates files (JSON, XML) into RDF using one of the mappings languages and examines the usability via a questionnaire for first-time users, which are have only basic skills in programming and linked data.
The results show that ShExML users tend to perform better than those of YARRRML and SPARQL-Generate. 
With SDM-RDFizer, RocketRML \cite{DBLP:conf/cikm/IglesiasJCCV20, DBLP:conf/esws/SimsekKF19}, there exist further concrete solutions that built upon RML.
However these approaches, as well as some of the ones mentioned before, follow the ETL approach, i.e., converting all data into a huge RDF file as opposed to on-demand query re-writing. 

\textit{xR2RML} is a language for mapping various types of databases to RDF which enables the translation of a broad scope of data sources via query re-writing \cite{DBLP:journals/tlsdkcs/MichelFM19}. 
To achieve this goal,
they propose a two-phase approach. First, they define an abstract query language derived from  SPARQL. 
Utilizing the xR2RML mapping language and leveraging R2RML-based SPARQL-to-SQL works, they introduce a generic method to translate a SPARQL graph pattern into their abstract query language. 
In the second phase, the abstract query is translated into the query language of a target database. 
For demonstration purpose, they apply the approach to MongoDB, but acknowledge that adopting a different NoSQL DBMS will still be challenging. 
This is mainly because these systems are generally optimized for fast storage and retrieval of vast collections of data, rather than on expressive query languages. 
The prototype implementation is open-source\footnote{\url{https://github.com/frmichel/morph-xr2rml/}} including the SPARQL-to-MongoDB translator and connectors for the MySQL and PostgreSQL relational databases.

Delva et al.\ \cite{DBLP:conf/esws/DelvaAHMD21} identify certain shortcomings of ShExML, xR2RML and combine concepts of both to introduce RMLFields. It extends RML with a nested iteration model empowering it to write nested loops over input data.
In another approach \cite{Araujo2017} called \textit{OntoMongo} the authors seek to delegate query execution to a NoSQL source engine. 
They rely on an object-oriented intermediate representation, where they map  from the ontology vocabulary to this object-oriented, conceptual layer. 
The aim is to simplify the mapping specification and make it independent of the underlying source database.

A team around the creators of \textit{Ontop} has been working on generalizing the OBDA concepts to non-relational data models. 
They have carried out experiments on MongoDB to extend the functionality of the system \cite{Botoeva2019}. 
They explain a two-step rewriting process of SPARQL queries into the MongoDB aggregate query language, where they represent the results of native queries as relations to provide a uniform view of non-relational queries. 
The idea is to view a collection of MongoDB documents in the nested relational model.

A similar implementation is reported in \cite{el2022virtual} aiming to fill the gap between NoSQL and the Semantic Web, in order to enable access to such databases and integration of non-relational data sources. 
They present the \textit{Ontop-CB} project which implements the query translation method based on the \textit{Ontop} system as a backbone, allowing to query Couchbase, a NoSQL document store. 
The key components are an OWL ontology, an access interface, mappings, a NoSQL database, a SPARQL to NoSQL query adjustment, and a JSON export. The approach exploits \textit{Ontop} answers to SPARQL queries by rewriting them into SQL queries and delegating their execution to the database. 
To do so, they also establish an object-oriented intermediate layer between the OWL ontology and the Couchbase data source.

The authors of \cite{Fathy2020} report the implementation of OBDA for a Neo4j graph database via two phases. 
First, the xR2RML mapping language is used to connect the RDF data model to the LPG data model.
In the LPG data model, edges may have multiple data properties, referred as `Composite Edge'. 
Then, relational graph algebra is used to translate a given SPARQL query into the Cypher query language native to Neo4j.

In line with the use of OBDA in NoSQL, the problem of ontology-mediated query answering over key-value stores is studied in \cite{mugnier2016ontology}. 
The authors create a rule-based language in which keys are used as unary predicates and rules are applied at the record stage (a record is a set of key-value pairs).

\textit{Chimera}’s main competitor is the Semantic ANalytics StAck (\textit{SANSA}) \cite{Lehmann2017}, that comes with \textit{Squerall}. 
\textit{Squerall} is a scalable OBDA engine that relies on the RML standard and allows querying several databases simultaneously.
\textit{Squerall} uses wrappers to query heterogeneous data sources directly in their original form. 
In the paper, they evaluate and compare five different data sources (Cassandra, MySQL, MongoDB, Parquet and CSV) and run queries involving multiple joins between the different sources.
They compare the \textit{Presto} and \textit{Apache Spark} processing engines for making \textit{SPARQL-to-SQL} conversions and find out that \textit{Presto} works faster in their specific evaluation. 
\textit{Squerall} addresses the variety challenge of Big Data in that it can conveniently be extended to embrace new data sources. 
This is the most immediate advantage over \textit{Chimera} and \textit{Ontop}, because those usually require a single DBMS (or \textit{Hive} store) in the backend. 
However, \textit{Squerall} does not comply to many W3C Recommendations. 
For example, basic SPARQL graph patterns OPTIONAL and UNION were not yet supported at the time of the first publication in 2019.
Similarly, it is unclear about how generic the framework truly is from a theoretical perspective beyond the evaluation scenario given in the paper. 
However, the work is a crucial step in the vision of implementing a polystore system that can query data in an on-the-fly-manner, i.e., no prior transformation.

\begin{table*}[htb]
  \centering
  \caption{Overview of OBDA techniques sorted by date of publication}
  \label{table-timeline-obda}
  \begin{tabular}{p{32mm}cp{36mm}p{42mm}p{44mm}}\hline
       & Year & Data Model  & Query Languages & Mapping Model \\ \hline
        \textit{Maestro} & 2011 & Relational  & SQL   & \textit{DL-Lite}  \\
        
        \textit{UltraWrap} & 2013   & Relational & SQL   & RDB2RDF  \\
        
        \textit{Optique} &  2013 & Relational  & SQL  &    RDB2RDF  \\
        
        
        \textit{MorphDB} & 2014 & Relational & SQL & R2RML   \\

        \textit{OnTop}  & 2014 & Relational & SQL & R2RML\\
        
        
        

        Mugnier et al.\ \cite{mugnier2016ontology} & 2017 & SQL \& Key-Value & SQL, XPath, JSONPath, MongoDB & NO-RL \\
        
        \textit{OntoMongo} & 2017 & Relational \& Document &  SQL \& MongoDB & Object-relational \& Object-Document  \\
        
        
        \textit{PolyWeb} & 2019 & Relational & SQL & R2RML \& RML \\
        
        \textit{OnTop} over MongoDB  & 2019 & Relational \& Document & SQL \& MongoDB & JSON-to-RDF \& SQL-to-RDF \\
        
        
        \textit{Ontario}  & 2019 & RDF \& Relational & SQL & RDF-MT\\
        
        \textit{Squerral} (\textit{SANSA})  & 2019 & Relational \& NoSQL  & Spark- \& Presto-SQL &  RML+FNO \\
        
        Fathy et al.\ \cite{Fathy2020}  & 2019 & labeled property graph & Cypher & xR2RML \\
        
        \textit{Obi-Wan} & 2020 & Relational \& Document  & SQL \& MongoDB  & (G)LAV view-based query rewriting \\
        
        \textit{OnTop4theWeb} & 2021 & REST (CSV, JSON, XML) & SPARQL & R2RML \\
        
        
        \textit{Chimera} & 2021 & Relational (Hive) & SparkSQL & R2RML  \\
        
        \textit{OntoCB} & 2021 & Document & Couchbase (N1QL) & Object-oriented \\
        
        \hline
    \end{tabular}
\end{table*}

\subsection{Discussion}
R2RML has been proven to be a useful mapping language for relational sources.
Simply extending the R2RML standard to support other types of sources does not necessarily carry on all its features \cite{Chortaras2018}.
For example, select conditions and transformation functions are supported by R2RML implicitly relying on the expressivity of SQL, but this cannot be applied for querying XML or JSON documents. 
In \cite{Corcho2020}, the authors argue first, typical mapping languages are designed to work with a specific data format, and second, they are designed in a format to be parsed by machines and they do not take into account human readability and compactness. 
YARRRML is clearly a contribution in that direction.
Because these languages are not necessarily interoperable, and many of them support a specific engine only, the next generation of OBDA systems would have to consider how to address translation mapping languages, in addition to the query re-writing techniques that have been widely addressed in the SOTA of OBDA so far (see \Cref{fig-timeline-obda}).

\textit{Squerall} extracts data, transfers it to an efficient format, and queries it in a scalable manner using \textit{Apache Spark}. 
In contrast, \textit{Ontop} predates the publication of the \textit{Chimera} and therefore does not utilize any Big Data technology. 
However, the comparison reported states that \textit{Ontop} is still competitive for the majority of the evaluated query types (12 in total), except for such queries that involve aggregating, where \textit{Squerall} is more efficient. 
This might come from the fact that \textit{OnTop} has more sophisticated query optimization techniques, while \textit{Squerall} relies solely on its processing engine. 
Furher, \textit{Squerall} does not support all query types.


Based on the existing literature we identify a historical development of OBDA techniques (see \Cref{fig-timeline-obda}) which is separated into four different periods:
\begin{figure}
    \centering
    \includegraphics[width=0.48\textwidth]{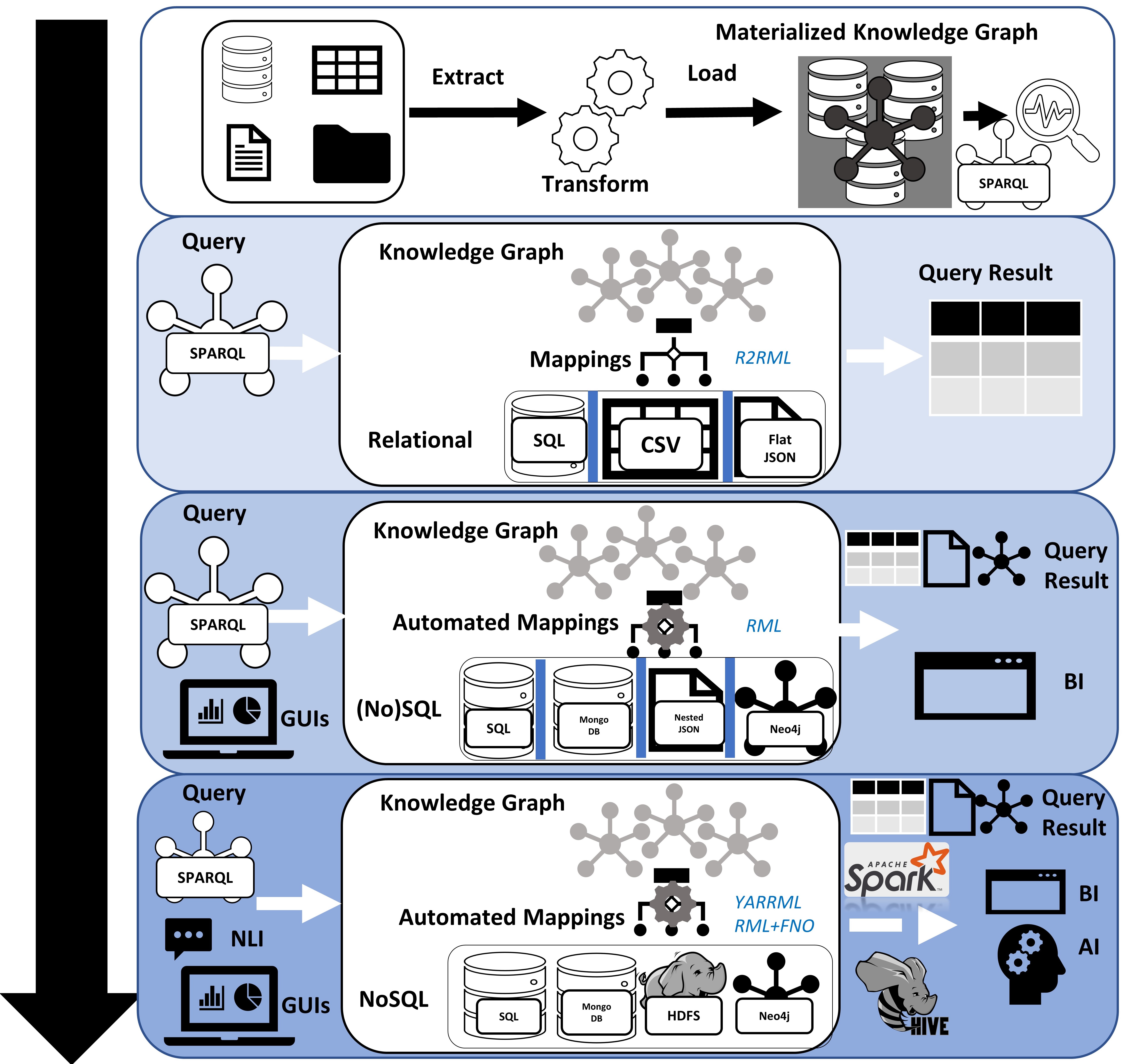}
    \caption{Timeline of OBDA techniques}
    \label{fig-timeline-obda}
\end{figure}
First, there are proposals following the ETL approach, i.e., transforming all data sources into a huge RDF graph to be queried using SPARQL. 
Although the approach might suggest a uniform query method for heterogeneous data sources, the question of data integration remains unanswered as data is only transformed into a common format, but it is unclear how they relate semantically to each other.
The approach also has the previously described deficiencies with regard to the liveliness of data and the overhead that comes with the initial transformation into RDF. 

These weak points were then addressed by the first platforms to perform OBDA over relational databases via query re-writing, which leaves the data in their respective DBMS retrieving it through the native via query optimization techniques.
Although these solutions display impressive progress, they are mostly limited to support only a single relational DBMS at once. 
The process of creating mappings is also mostly manual and hence tedious, error-prone, and tailored towards a single static ontology. 

The third period bridges the gap to access NoSQL databases and more flexible data models such as nested structures, graphs, or REST APIs and also expands the query result space to NoSQL data models.
In this period many 'single-purpose' solutions arise that extend support to a single NoSQL DBMS or to semi-structured files like JSON and XML, but no solution exists that is flexible enough to query arbitrary databases. 
In this period we also observe the first real-world applications that built GUIs for these methods to perform business intelligence effectively, e.g., \cite{Kharlamov2017a}.
In parallel, the development of algorithms for automated generation of mappings gain increasing momentum, some of which have been discussed in \cref{chapter-automatic}.
Early proposals with a focus on OBDA include the already mentioned \textit{BootOX} complemented by \textit{MIRROR} \cite{Medeiros2015}, both promoting the automatic generation of W3C R2RML from RDBMSs.
They are followed by Hazber et al.\cite{Hazber2016}, Heyvaert et al.\ \cite{Heyvaert2017}, \textit{AutoMap4OBDA} \cite{Sicilia2016}, \textit{MILAN} \cite{Mathur2018} and Iglesias-Molina et al.\ \cite{IglesiasMolina2020}  (\textit{Mapeathor}\footnote{\url{https://morph.oeg.fi.upm.es/tool/mapeathor/swagger/$\#$/default/post\_tool\_mapeathor\_api\_}}) each presenting different algorithms to improve the mapping quality, decreasing the task complexity, and hence the user effort for creating R2RML mappings.

Finally, the most recent approaches add efficient, scalable, and distributed processing engines like \textit{Apache Spark} or \textit{Hive} and storages like \textit{HDFS}. The \textit{Chimera Suite} supports all W3C recommendations for the R2RML mapping language to translate SPARQL to SQL and bridges an important gap between semantic and Big Data technologies. 
However, at the same time, it is limited to \textit{Hive} and the relational data model. 
Furthermore, the traditional GAV-based mappings have the drawback that one needs to revisit them if schema changes happen at the source level. 
For example, the \textit{Ontop} docker container\footnote{\url{https://github.com/chimera-suite/use-case}} needs to be restarted every time changes are made to the mapping definitions or the data sources. 
This problem is partly lifted by utilizing GLAV mappings, but they have not gained widespread attention yet and scalable open-source proposals are to be presented.
On the other side \textit{Squerall} supports arbitrary NoSQL databases while also utilizing scalable processing engines, but does not support the sophisticated SPARQL queries\footnote{\url{https://github.com/EIS-Bonn/Squerall/wiki/Squerall-Basics\#3-sparql-query-interface}}.
Furthermore, we observe the emergence of data-driven methods, in particular scaling usability of Machine Learning (ML) analytics, which requires managing a large variety of heterogeneous data with a unified mechanism, in order to save time on developing new ML solutions, and to reuse already developed ones.
For this, companies commit to the development of a long-term common semantic frameworks for industrial data management and ML analysis \cite{Zhou2021}.

%% file: chapters/applications.tex
\section{Applications}\label{chapter-applications}
The closer collaboration between human-machine and machine-machine systems has revolutionized the current industrial landscape, leading to Industry 4.0 (also known as Industrial Internet-of-Things, IIoT) \cite{ustundag2018industry}.
The key drivers of IIoT include the continuous addition of new types of devices with their own data models.
This data explosion requires highly agile data models that enable monitoring, processing, optimizing, and analyzing data to derive insights for better decision-making.
However, effective data utilization demands data integration, encompassing cleaning, de-duplication, and semantic homogenization.
We proceed by presenting two hand-selected use cases where semantic data management has transcended the realm of academia and has been applied to real-world IIoT problems.

\subsection{Manufacturing}
The company Bosch has recognized the potential of knowledge graphs, and a series of subsequent publications highlight the company's significant advancements in the field of semantic data management and integration.
Firstly, OBDA using the \textit{SANSA Stack} as well as \textit{OnTop} has been reported in the area of Surface Mount Technology (SMT), the process for mounting electronic components on printed-circuit boards \cite{Mami2020, DBLP:conf/semweb/KalayciGLXMKC20}.
This process involves several subprocesses executed by specialized machines, each generating substantial amounts of data. These data contain semantic interoperability conflicts, such as variations in object naming. To effectively explore this data, these conflicts need to be resolved by mapping them to ontology terms.
The development of the SMT Ontology resulted from a collaborative effort spanning several weeks, involving a line engineer, two line managers, an SMT process expert, an SMT data manager, a project manager, two Big Data managers, and four semantic experts.
Thus, adopting the VKG approach, which focuses on producing high-quality mappings, is a labor-intensive process requiring the combination of three types of knowledge initially: domain knowledge, database schema details, and understanding of the VKG approach.
However, once achieved, an integrated Data View provided by the SMT Ontology enables domain experts to perform intricate product analysis tasks on previously unintegrated raw data.

The development of knowledge graphs is a major undertaking for the company, involving substantial initial effort but promising long-term solutions for various complex challenges.
Firstly, it serves the primary purpose of storing knowledge and supporting material science engineers in their information search \cite{Stroetgen2019}.
Additionally, the focus extends to data-driven methods, particularly enhancing the usability of Machine Learning (ML) analytics. This entails managing heterogeneous data from a wide array of sources through a unified mechanism, saving time in developing new ML solutions, and reusing previously developed ones \cite{Zhou2021a}.
Their system comprises several semantic artifacts: a core ontology that encapsulates general knowledge of the manufacturing process, domain ontologies reflecting domain-specific knowledge, and \textit{data-to-domain ontology} mappings that link raw data attributes to domain ontology terms to enable OBDA.
Several semantic modules support the management of the three fundamental building blocks: the KG generation module, a mapping reasoner/annotator, and a data integration module.
The company's primary goal is to leverage ML analytics to achieve a high degree of production automation. All these components serve as prerequisites, culminating in a platform known as \textit{SemML} (Semantic Machine Learning) \cite{DBLP:conf/semweb/KalayciGLXMKC20, Zhou2020}, addressing three core challenges through semantically enhanced ML:
(1) communication, as the workflow involves collaboration among experts from diverse areas, including data scientists, engineers, process experts, and managers with distinct backgrounds, making communication time-consuming and error-prone;
(2) data integration; and
(3) generalizability of ML models.

\subsection{Smart City}
The digital transformation encompasses not only industrial applications but also infrastructures for services of general interest, such as transportation networks and associated transport options, water and energy supply, as well as waste and wastewater disposal. These activities are commonly associated with the term \emph{Smart City}. In a Smart City, intelligent information and communication technology is employed to enhance participation, quality of life, and to create an economically, ecologically, and socially sustainable community or region.

As described in \Cref{chapter-compare-datalakes}, Bagozi et al.~\cite{Bagozi2019} and Bianchini et al.~\cite{Bianchini2018, Bianchini2020} utilize indicators to enable personalized data lake exploration, demonstrating the model's effectiveness in the Smart City environment due to its data and user role diversity.
Furthermore, Bianchini et al.~\cite{bianchini-smart-city} present a tool that implements a semantic layer over a heterogeneous ecosystem of data sources, serving as a starting point for semantics-enabled applications or data source exploration. Their pipeline involves three steps: (1) Lexical enrichment of attributes, standardizing concept names for annotation; (2) semantic annotation of data attributes; and (3) creation of semantic relationships.

Similarly, Pomp et al. introduce a \textit{Semantic Data Marketplace for Easy Data Sharing within a Smart City}, applying a predecessor of \textit{PLASMA} (refer to \Cref{chapter-OBDM}) to the Smart City use case. The semantic data marketplace was employed in a project\footnote{\url{https://youtu.be/3G9IQgK09sU}}, requiring teams to develop Smart City applications based on datasets provided by city departments from three different cities.
Both approaches involve a Data Producer and a Data Consumer. The Data Producer, a domain expert, is responsible for uploading datasets and enriching them with semantic metadata from knowledge graphs.
Using the knowledge graph and associated semantic models, data consumers can find and comprehend data by searching and reviewing semantic concepts.
VC-SLAM \cite{Burgdorf2022} (refer to \Cref{chapter-semantic-model-generation}), composed of 101 datasets with corresponding semantic labels and a semantic model, has been constructed using Smart City data extracted from Open Data portals.
A recent survey \cite{nahhas-smartcity} introduces semantic Internet-of-Things technologies, which may play a crucial role in addressing essential Smart City issues, primarily interoperability challenges.

%% file: chapters/challenges.tex
\section{Challenges}
\label{sec:challenges}
\begin{figure}
    \centering
    \includegraphics[width=0.8\columnwidth]{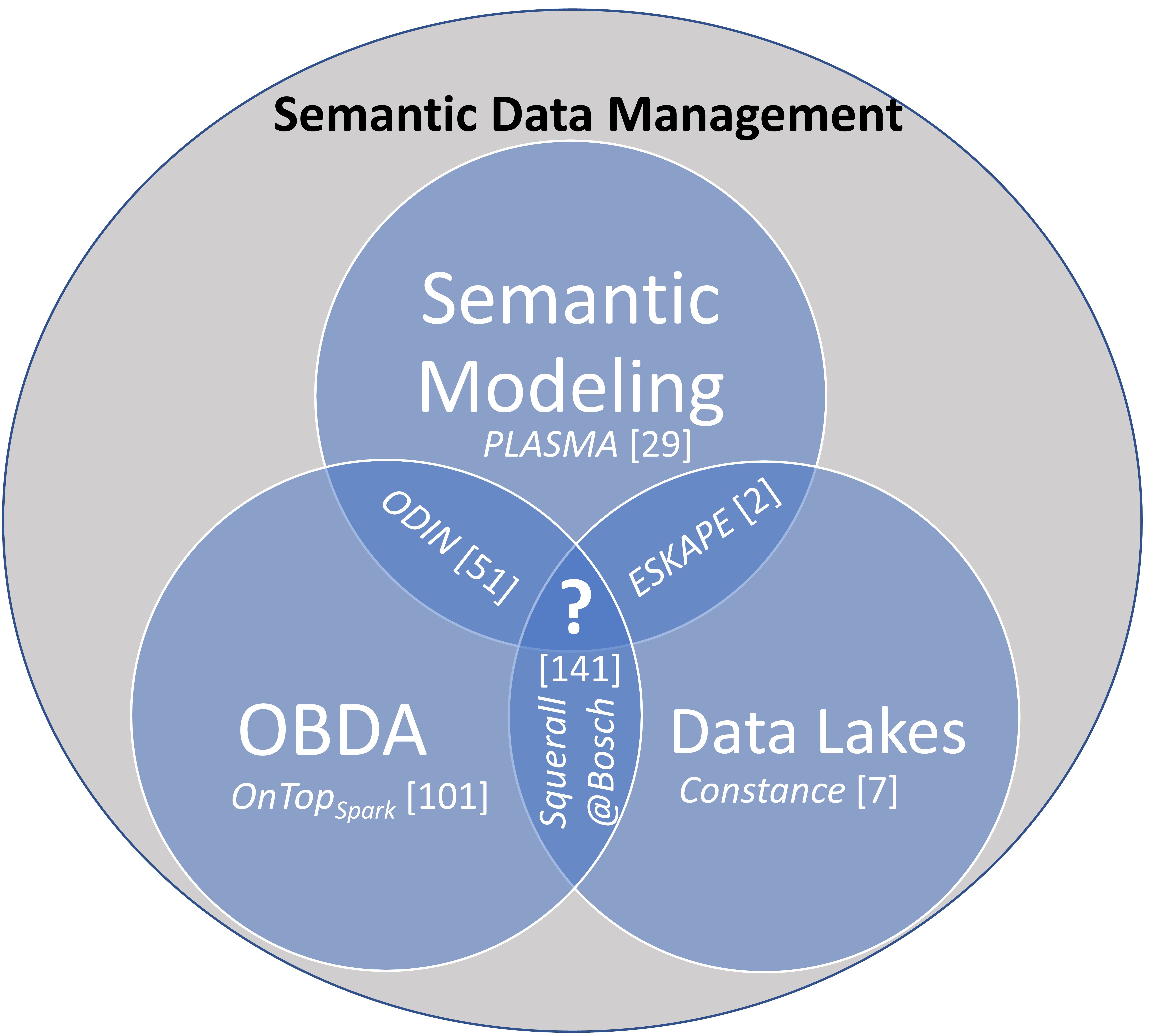}
    \caption{Venn-Diagram to illustrate our view of the current state in the field of Semantic Data Management}
    \label{fig-venn}
\end{figure}

Even though we have seen various tools in which KGs play a crucial role, 
a comprehensive solution that provides truly scalable access to heterogeneous data based
on ontology-based mappings, including a semantic modeling component to create and maintain the mappings, is not yet available.
We have illustrated this gap in \Cref{fig-venn}.
In the Venn-diagram, we put three topics in relation: OBDA, semantic modeling, and data lakes. 
Semantic Data Management may include more approaches, such as the ETL approaches neglected here.
The intersection between semantic modeling and OBDA is represented for example by \textit{ODIN}.
The combination of OBDA for Big Data and data lakes is illustrated for example by the manufacturing use case \cite{Mami2020}.
However, it involved a great amount of manual labour by experts, is not available (neither open-source nor as commercial system) and we found no evidence that the semantic modeling techniques discussed in \cref{chapter-automatic} are applied in this use case. 

Semantic modeling is enabled within a data lake by \textit{ESKAPE}.
However, a custom data format named Semantic Linked Tree (SLT) is used, which is a JSON-based data format in which the semantic labels are attached to the raw data values.
This transformation can be interpreted as \textit{semantic lifting}.
Due to its rather early development, the system also does not incorporate scalable data storage and processing engines inherent to most modern data lakes.

OBDA has grown from a long tradition, it has been reported in production several times, and recent proposals, such as \textit{SANSA} and \textit{Chimera} display meaningful progress towards semantics-based on-demand data access.
However, the creation of the required mappings is a tedious, manual task for the user. There is not yet enough assistance either by good user interfaces or automatic procedures.

The field faces several challenges today:
\begin{enumerate}[leftmargin=0.5cm]
    \item \textit{Initial overhead \& usability}: 
    Creating meaningful semantic models for a large number of heterogeneous data sets comes with a huge initial overhead in generating KGs and the corresponding connections. 
    First, generating domain-specific conceptualizations is already time-consuming and resource-intensive, quite apart from the fact that this task is truly an entire field of its own.
    Even if we assume to have a versatile set of suitable KGs at hand, the creation of suitable mappings to data sources is still a time-consuming process. 
    As presented in this survey, there are some proposals for automatic creation of semantic models and mappings; yet, more emphasis should be put on the required human input (see \textit{Technical abstraction}). 

    \item \textit{Evaluation}:
    Automated generation of semantic labels and models has gained much attention lately; however, it is unclear how accurate the methods really are, beyond the scope of the test sets used in the particular publication.
    For semantic labeling, \textit{SemTab} has presented various data sets and benchmarks and research is under way for more sophisticated table interpretation.
    For semantic models, \textit{VC-SLAM} is the very first proposal towards standardization of a set of ontologies and data sets alongside the corresponding semantic models.
    The emergence of annual challenges and the publication of benchmarks will be helpful and could be further streamlined by an initiative similar to the OAEI\footnote{\url{http://oaei.ontologymatching.org/}} for ontology alignment.
    
    \item \textit{Technical interoperability}:
    Considering the variety of today's data landscape, one has to point out that the majority of methods for semantic labeling and modeling are based on tabular data. The field is yet far away from covering all NoSQL data models uniformly.
    This is also the focus of modern OBDA research. 
    While scalable data access via ontologies can be regarded as solved to a large extent for relational data, the focus has shifted more towards incorporating different NoSQL query languages as well as supporting federated query processing against multiple databases, file systems and other miscellaneous data sources simultaneously. 
    Any such solution \textit{must} be compatible with W3C standards for the Semantic Web to ensure maximal interoperability between systems.
    
    \item \textit{Technical abstraction}: 
    The overall quality of a semantic model must ultimately be refined by the human operator.
    This human-in-the-loop process needs guidance for unfamiliar users through a strong technical abstraction, i.e., technical details must be hidden from the user.
    It is hard to imagine that even with very enhanced AI techniques (see below \textit{Leverage AI}), human verification and refinement of mappings will become obsolete. Therefore, enhanced user interfaces that focus on usability also for non-technical users are required in this domain.

    \item \textit{Applicability for Big Data}: 
    Solutions like \textit{Squerall} and \textit{Chimera} are promising approaches to address OBDA also in Big Data scenarios. 
    Yet, there are many limitations in the expressiveness of queries or mappings. 
    Furthermore, they are bound to specific versions of Big Data platforms which complicates their deployment. 
    The maturity of such solutions could be improved if a larger community would continue the development of these prototypes, in order to remove the aforementioned limitations and make them applicable to a wide variety of NoSQL / Big Data systems.
    
    \item \textit{Leverage AI}:
    Since the publication of ChatGPT, it has been discussed for many human tasks whether they can be solved by Large Language Models (LLMs) in the future. 
    This is also relevant for all tasks related to data integration as ChatGPT can describe data sources, create ontologies, find relationships between different data sets, to name just a few examples of what can be achieved with LLMs. 
    In view of the latest development of AI techniques, we have to expect that these systems will be able to solve more complex data integration tasks in the future of which the first papers are appearing already \cite{korini2023column}.
    It would be interesting to see how such general-purpose AI systems based can be customized and optimized for specific data integration tasks within data lake systems at scale. 
    On the other hand, we need to examine the processes discussed in this survey to see at which points and how an AI system based on LLMs can be integrated to improve the overall process.
    
\end{enumerate}

%% file: chapters/concl.tex
\section{Conclusion}
In this article, we have given an overview of semantics-based methods for data management, access, and integration and related those findings to current semantic data lake proposals. 
Conclusively, we can state that the community faces several challenges and a gap in today's landscape between present data lake platforms, OBDA and semantic technologies for modeling the context in which heterogeneous data sets arise. 
The heterogeneity inherent to Big Data applications is still a concern in the current OBDA platforms although there has been significant progress recently.

In the previous section, we have pointed out several challenges that can be addressed in this research area. We are confident that Big Data and Semantic Web technologies can benefit from each other and that more enhanced solutions for semantic data lakes will become available in the next years.